\documentclass[a4paper,11pt]{article}
\pdfoutput=1 

\usepackage{jinstpub} 
\usepackage{color}
\usepackage[normalem]{ulem}

\usepackage{lineno}

\usepackage{graphicx}
\usepackage{subfigure}

\usepackage{siunitx}

\DeclareSIUnit \mm {\milli\meter}
\DeclareSIUnit \cm {\centi\meter}
\DeclareSIUnit \us {\micro\second}
\DeclareSIUnit \ms {\milli\second}
\DeclareSIUnit \C {\coulomb}
\DeclareSIUnit \pA {\pico\ampere}
\DeclareSIUnit \pC {\pico\coulomb}
\DeclareSIUnit \fC {\femto\coulomb}
\DeclareSIUnit \fF {\femto\farad}
\DeclareSIUnit \pF {\pico\farad}
\DeclareSIUnit \nF {\nano\farad}
\DeclareSIUnit \mV {\milli\volt}
\DeclareSIUnit \kV {\kilo\volt}
\DeclareSIUnit \V {\volt}
\DeclareSIUnit \GOhm {\giga\ohm}
\DeclareSIUnit \MOhm {\mega\ohm}
\DeclareSIUnit \ton {\tonne}
\DeclareSIUnit \kton {\kilo\tonne}
\DeclareSIUnit \kt {\kilo\tonne}
\DeclareSIUnit \Mt {\mega\tonne}
\DeclareSIUnit \eV {\electronvolt}
\DeclareSIUnit \keV {\kilo\electronvolt}
\DeclareSIUnit \MeV {\mega\electronvolt}
\DeclareSIUnit \GeV {\giga\electronvolt}
\DeclareSIUnit \km {\kilo\meter}
\DeclareSIUnit \kW {\kilo\watt}
\DeclareSIUnit \MW {\mega\watt}
\DeclareSIUnit \Hz {\hertz}
\DeclareSIUnit \MHz {\mega\hertz}
\DeclareSIUnit \kHz {\kilo\hertz}
\DeclareSIUnit \mrad {\milli\radian}
\DeclareSIUnit \year {year}
\DeclareSIUnit \POT {POT}
\DeclareSIUnit \sig {$\sigma$}
\DeclareSIUnit\parsec{pc}
\DeclareSIUnit\lightyear{ly}
\DeclareSIUnit\foot{ft}
\DeclareSIUnit\ft{ft}

\usepackage{newtxmath,newtxtext}

\newcommand{\fixme}[1]{}

\newlength{\figwidth}
\newlength{\fighalfwidth}
\collaboration{The MicroBooNE Collaboration}

\title{Noise Characterization and Filtering in the MicroBooNE Liquid Argon TPC}

\author[g]{R.~Acciarri}
\author[bb,1]{C.~Adams\note{now at: Harvard University, Cambridge, MA, 02138, USA}}
\author[h]{R.~An}
\author[c]{J.~Anthony}
\author[y]{J.~Asaadi}
\author[a]{M.~Auger}
\author[g]{L.~Bagby}
\author[bb]{S.~Balasubramanian}
\author[g]{B.~Baller}
\author[n]{C.~Barnes}
% ...

\affiliation[a]{Universit{\"a}t Bern, Bern CH-3012, Switzerland}
% ...
\affiliation[c]{University of Cambridge, Cambridge CB3 0HE, United Kingdom}
% ...
\affiliation[g]{Fermi National Accelerator Laboratory (FNAL), Batavia, IL 60510, USA}
\affiliation[h]{Illinois Institute of Technology (IIT), Chicago, IL 60616, USA}
% ...
\affiliation[n]{University of Michigan, Ann Arbor, MI, 48109, USA}
% ...
\affiliation[y]{University of Texas, Arlington, TX, 76019, USA}
% ...
\affiliation[bb]{Yale University, New Haven, CT, 06520, USA}

\date{}
\abstract{
The low-noise operation of readout electronics in a liquid argon time projection chamber
  (LArTPC) is critical to properly extract the distribution of
  ionization charge deposited on the wire planes of the TPC, 
  especially for the induction planes.
  This paper describes the characteristics and mitigation of the observed noise in
  the MicroBooNE detector. The MicroBooNE's single-phase LArTPC comprises two induction planes 
  and one collection sense wire plane with a total of \num{8256} wires.
  Current induced on each TPC wire is amplified and shaped by 
  custom low-power, low-noise ASICs immersed in the liquid argon.
  The digitization of the signal waveform occurs outside the cryostat.
  Using data from the first year of MicroBooNE
  operations, several excess noise sources in the TPC were
  identified and mitigated.
The residual equivalent noise charge (ENC) after noise
  filtering varies with wire length and is found to be below 400
  electrons for the longest wires (\SI{4.7}{\meter}). The response is consistent
  with the cold electronics design expectations and is found to be
  stable with time and uniform over the functioning 
  channels. This noise level is significantly
  lower than previous experiments utilizing warm front-end
  electronics.
  }


\begin{document}
\maketitle
\flushbottom

\section{Introduction}\label{sec:intro}

The liquid argon time projection chamber
(LArTPC)~\cite{lartpc_1,lartpc_2,lartpc_3,lartpc_4} is a totally
active three-dimensional (3D) tracking calorimeter that enables the detection of
accelerator neutrinos, supernova burst neutrinos, and sensitive searches for proton decay, 
with unprecedented spatial resolution for detectors of this scale.

The MicroBooNE detector, a 170 ton LArTPC, commenced operations in the Booster Neutrino
Beam in October 2015. During an initial operating period from October 2015 through July 
2016, several distinct types of TPC noise were observed in the
MicroBooNE data which exceeded expectations for the
types of noise inherent to the electronics.
Characteristics of the noise in the frequency and time domains were
determined, and tests to identify the possible sources of the excess
noise were performed. An offline noise filter was developed that
eliminates most of the excess noise while retaining excellent signal preservation. 
The filtered noise levels are consistent with expected irreducible
noise inherent to the cold front end application-specific integrated circuits (ASICs) and the analog-to-digital converters (ADCs).

In summer of 2016, several hardware upgrades were performed to mitigate the
two largest sources of excess noise. After the successful upgrade, the residual excess 
noise was largely diminished reducing the requirement for offline filtering.

Details are presented in this paper as follows: section~\ref{sec:RC}
describes the basic design features and the inherent noise of the
readout electronics. Section~\ref{sec:non-func} describes the status of
the TPC readout after installation.  The details of the observed
excess TPC noise in MicroBooNE, the corresponding offline noise filter, and
impact of the noise filter on the real signals induced by
ionization charge are described in section~\ref{sec:types}.
Section~\ref{sec:RMS} describes the residual noise level after the
noise filter was implemented and section~\ref{sec:PSNR} presents a
metric to compare peak-signal from minimum ionizing tracks observed in
data with the measured noise. Hardware upgrades to mitigate the two
largest sources of excess noise and their effect are summarized in
section~\ref{sec:hardware}.  The conclusions and outlook are summarized
in section~\ref{sec:summary}.

\section{MicroBooNE detector and readout electronics}\label{sec:detector}

The MicroBooNE
detector~\cite{Acciarri:2016smi} is a recently built LArTPC designed
to observe interactions of neutrinos from the on-axis Booster~\cite{AguilarArevalo:2008yp} 
and off-axis NuMI~\cite{Adamson:2015dkw} neutrino beams at the Fermi National Accelerator Laboratory in
Batavia, IL. The detector consists of a 
$ \SI{2.56}{\meter} \times \SI{2.32}{\meter} \times \SI{10.36}{\meter}$ 
TPC for charge detection, and an array of 32 photomultiplier tubes (PMTs)~\cite{pmt}
that detect scintillation light for triggering, timing, and reconstruction 
purposes. The active mass of the detector is 85 metric tons of
liquid argon (LAr). The TPC is housed in a foam-insulated evacuable cryostat vessel
with a \SI{2.56}{\meter} drift distance bounded on one end by the cathode plane
and the other by three parallel anode wire readout planes.
In the
drift direction, these planes are labeled ``U'', ``V'', and ``Y''.
The U, V, and Y planes contain \num{2400}, \num{2400}, and \num{3256} wires, respectively.
The wire spacing within a plane is \SI{3}{\mm}, and the 
planes are spaced by \SI{3}{\mm}.
The wires comprising the Y plane run vertically and the
wires in the U and V planes are oriented $\pm \SI{60}{\degree}$ with
respect to vertical. While the TPC cathode is designed to operate at a high voltage (HV) of 
\SI{-128}{\kV}, corresponding to a drift electric field of \SI{500}{\V/\cm},
MicroBooNE currently operates at \SI{-70}{\kV}, corresponding to a drift field of \SI{273}{\V/\cm}.  
In this field, the ionization electrons drift at a speed of \SI{1.114}{\mm/\us}~\cite{yichen}. 

\begin{figure}[!h!tbp]
\includegraphics[width=\figwidth]{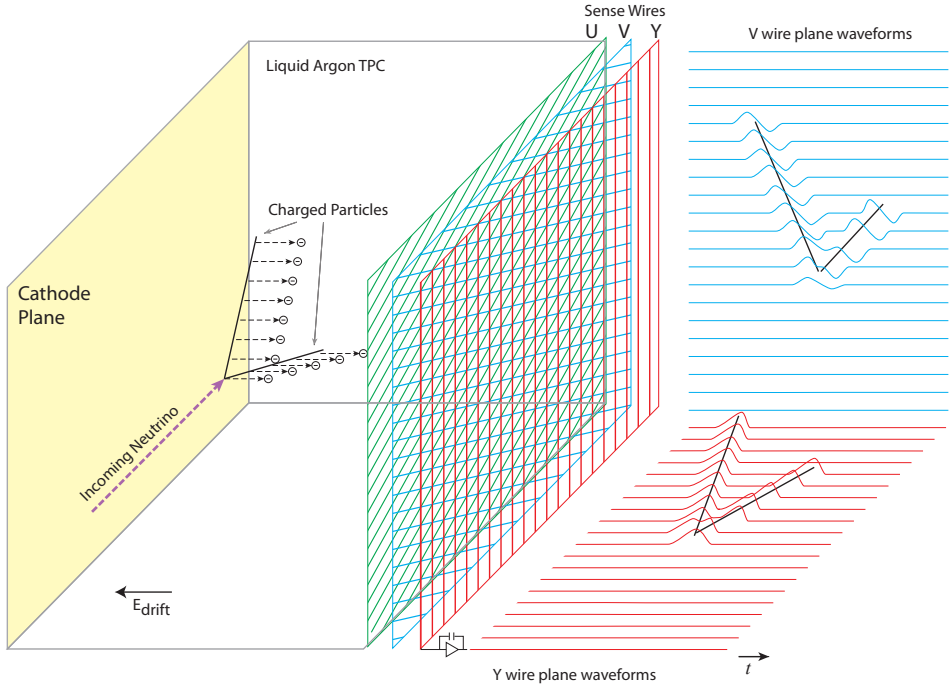}
\caption[TPC Basics]{Diagram illustrating the signal formation in a
  LArTPC with three wire planes~\cite{Acciarri:2016smi}. For simplicity, the signal in the
  first U induction plane is omitted in the illustration. Planes are positioned in the order U, V, Y with 
  the Y plane being farthest from the cathode plane.}
\label{tpccartoon}
\end{figure}

Figure~\ref{tpccartoon} illustrates the signal formation in
the MicroBooNE TPC.  The ionization electrons produced along an
energetic charged particle track drift through the LAr along the
electric field lines toward the anode wire planes.
Bipolar currents are induced on the wires of the U and V planes,
commonly referred to as the \textit{induction planes}, as the charge
drifts towards and then past them.
A unipolar signal is induced on a wire of the Y plane or
\textit{collection plane}, as all nearby ionization is collected.
In addition, there are second order effects due to long range
induction that contribute to the induced current waveform on any given
wire~\cite{Shockley,Ramo}.
Bias voltages for the U, V and Y planes, \SI{-110}{\V}, \SI{0}{\V}, and
\SI{+230}{\V}, respectively, are applied to each wire plane to ensure
the two induction planes satisfy the transparency
condition~\cite{Bunemann_Cranshaw_Harvey_1949} that all drifting
electrons pass by the induction plane wires and fully collect on wires
in the last (collection) plane.

\begin{figure}[!h!tbp]
  \center
  \includegraphics[width=\figwidth]{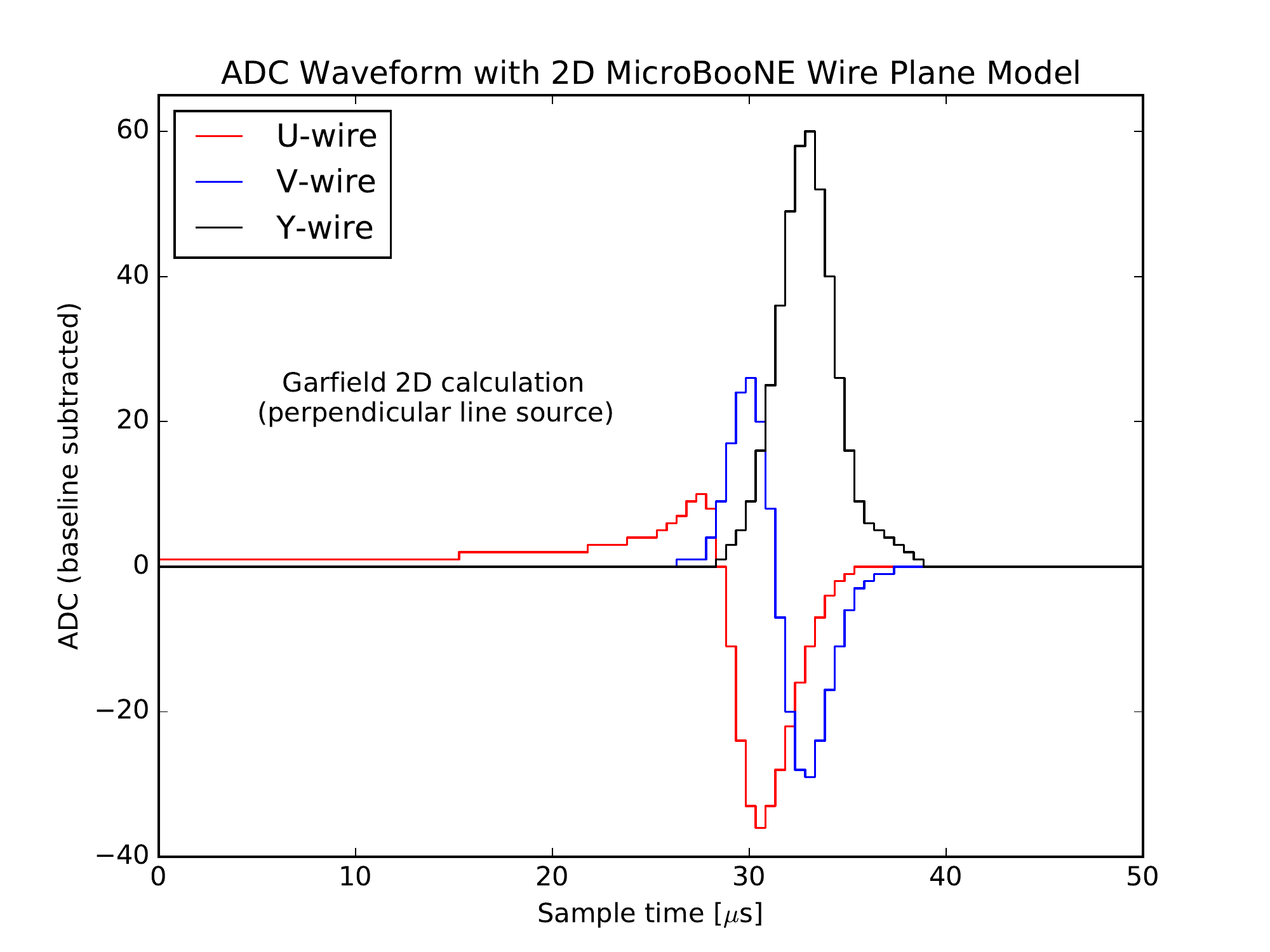}
  \caption[TPC Signal]{The digitized signals from a central wire from
    each plane that are induced by an ideal MIP track in a 2D model
    of the MicroBooNE TPC. The high resolution field response functions used
    were calculated with Garfield 2D~\cite{Veenhof:1998tt}.  They were defined on drift
    paths spaced by 10\% of the pitch which cover ten wires to
    either side of the three central wires.  An electronics response
    function with an amplifier gain of \SI{14}{\mV/\fC} and peaking
    time of \SI{2}{\us} is used to model the ASIC and an additional
    gain of 1.2 is used to account for subsequent amplification
    stages.  The track is an ideal line source that runs perpendicular
    to all wires, consists of 5300 ionization electrons per mm and
    spans the transverse domain of the calculation.}
  \label{fig:tpc_signal}
\end{figure}

Figure~\ref{fig:tpc_signal} illustrates the digitized simulated signal waveforms on 
the induction and collection planes assuming an ideal minimum ionizing
particle (MIP) track.  
Since the ionization electrons are collected by the 
Y plane wires, integrating the waveform recorded from each Y wire gives 
a measure of the charge deposited along that section of the track.
The peaks of the bipolar signals on the U and V induction wires are typically 
a factor of \numrange{2}{3} smaller than the peak of the unipolar signal on the
Y collection wires for this kind of signal.


The requirements on the TPC readout electronics performance are driven
by the physics goals of MicroBooNE. The dynamic range and noise
performance of the TPC readout are determined by the need to clearly
measure the ionization charge arriving at the wire planes for a range
of activity varying from the small charge created by a MIP to the large ionization deposited by stopping protons
emerging from the breakup of an argon nucleus~\cite{uboone_2}. 

The induced current on each wire is amplified and shaped through one dedicated
channel (out of 16) of a custom designed complementary
metal-oxide-semiconductor (CMOS) analog front-end cold
ASICs~\cite{Radeka:2011zz}.
The analog front-end ASICs, which include a pre-amplifier, shaper, and 
signal driver are located inside the cryostat along with 
the wire bias voltage distribution system, decoupling capacitors, 
and calibration networks. HV capacitors of \SI{2.2}{\nF} rated to \SI{1}{\kV} and
\SI{4.7}{\nF} rated to \SI{630}{\V} are mounted on the wire carrier base boards~\cite{Acciarri:2016ugk} to AC-couple 
the Y and U wires to the readout electronics. Current-limiting resistors of \SI{22}{\MOhm} 
are placed between the bias voltage bus and each wire in case a wire is shorted to a different wire plane.
V wires do not have HV bias resistors or decoupling capacitors, as they 
are DC coupled to the readout electronics. 
A set of ASICs is mounted on one front-end motherboard (FEMB),
which is placed
close to the end of the sense wires, embedded inside the LAr in order to
minimize the capacitive load - and hence noise - into the ASICs.  As will be 
discussed in section~\ref{sec:RC}, the cold temperature further reduces
the noise with respect to the room temperature operation of these
ASICs.

The ASICs can operate at one of four gain settings
(\SIlist{4.7;7.8;14;25}{\mV/\fC}). The gain determines the peak
height of the output voltage waveform because of the collection of an
impulse of unit charge. Four peaking time settings
(\SIlist{0.5;1.0;2.0;3.0}{\us}) are available. The peaking time is
defined as the time difference between 5\% of the peak position and 
the signal peak of an impulse response. The 
ASICs are nominally operated at \SI{14}{\mV/\fC} gain and \SI{2}{\us}
peaking time settings throughout data taking.

The output dynamic range of the
cold ASIC is \SI{1.6}{\V}.
The output signals from the ASICs are transmitted over of distance of 
\SIrange{2.5}{5.5}{\meter} by
twisted-pair copper cable, through a warm flange, and to an
intermediate amplifier line-driver which resides in an enclosure that forms a Faraday cage with the cryostat.
The intermediate amplifiers compensate for losses incurred by the
signals as they are then driven over \SI{10}{\meter} of
shielded, twisted-pair cables to the data acquisition (DAQ) TPC readout boards 
in crates on the platform above the detector.
At this point, these signals having picked up a gain of \num{1.2} 
are digitized by AD9222~\cite{AD9222} ADCs.

The ADC has a dynamic range of \SI{2}{\V}, digitizes with a
sampling rate of \SI{2}{\MHz}, and records \SI{12}{bit} samples with
an inherent ADC noise that gives an effective number of bits of
\num{11.3}~\cite{enob}.
In order to match the output
voltage range of the cold ASIC and the input range of the ADC, a
resistor-capacitor (RC) circuit with \SI{1}{\ms} time constant is
included in the intermediate amplifier to remove the ASIC baseline
voltage. The intermediate amplifier and receiver/ADC board provides the net gain of \num{1.2} to the
signal helps match the differing voltage ranges of the cold ASIC and ADC.
A second RC circuit with a \SI{1}{\ms} time constant is installed in
front of the ADC to make the ADC pedestal\footnote{The pedestal
  value of the ADC output is an effective offset. It is defined by the
  integration of the excess signal from the gate pulse which allows
  current to pass into the ADC. } independent of the intermediate
amplifier baseline.

\begin{figure}[!h!tbp]
\center
\includegraphics[height=\figwidth,angle=-90]{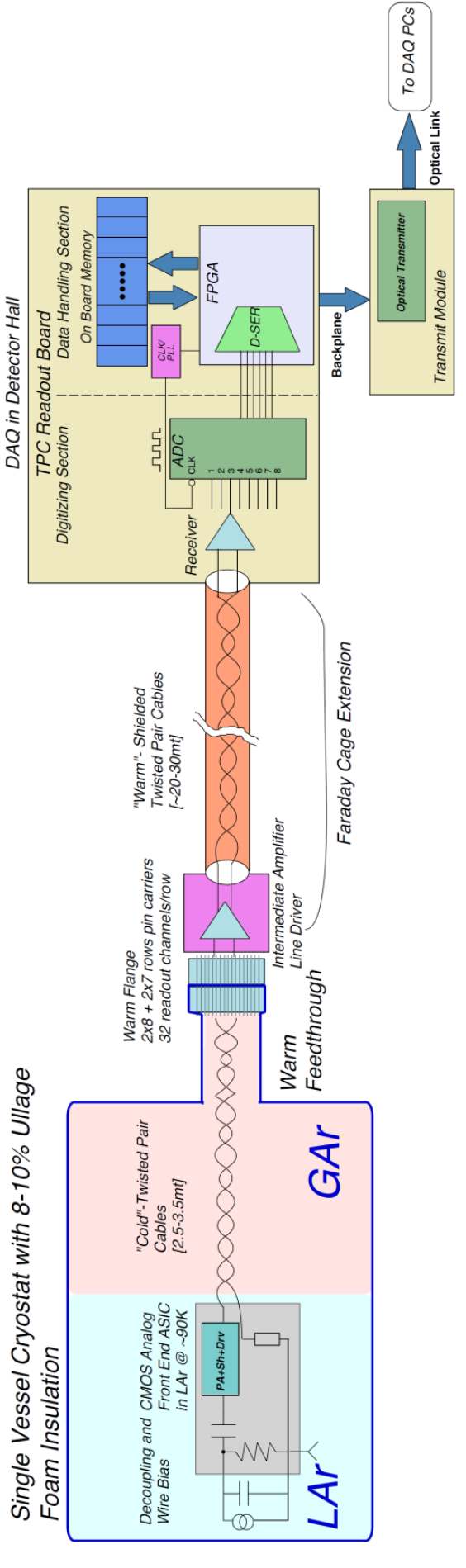}
\caption[Readout Electronics Schematic]{Schematic of MicroBooNE cold and 
warm electronics readout chain~\cite{uboone_2}. See text for more 
explanation.}
\label{elecschema}
\end{figure}

Each DAQ readout window is \SI{4.8}{\ms} in duration, which corresponds
to \num{9600} recorded samples (also called time ticks or ticks), and
is slightly more than twice the time needed for an ionization electron to
drift the full width of the detector. The digitized signal from the
ADC module is passed directly to a field-programmable gate array
(FPGA) chip for initial data processing, data reduction, and
preparation for readout by the DAQ system.  A schematic of the
MicroBooNE readout electronics is shown in figure~\ref{elecschema}.

\section{Front-end readout electronics response and inherent noise}
\label{sec:RC}

To better understand the overall electronics response and the inherent
noise of the MicroBooNE readout electronics, the entire front-end
electronics chain up to the ADC can be described in terms of four basic
circuits:
\begin{enumerate}
\item A charge pre-amplifier in the cold ASIC.
\item A subsequent shaping circuit in the cold ASIC.
\item An RC circuit in the intermediate, line-driver amplifier.
\item A pedestal-adjusting RC circuit located just before the ADC input.
\end{enumerate}

\subsection{Impulse response}
\label{sec:impulse}

As described earlier, the analog signal from the MicroBooNE front-end
electronics is digitized by a \SI{12}{bit} ADC 
with a sampling rate of \SI{2}{\MHz}.
According to the Nyquist theorem~\cite{PhysRev.32.110}, the
information in a continuous signal is fully sampled if the signal
contains no power at frequencies above the Nyquist frequency of
one-half the sampling rate.
The MicroBooNE
pre-amplifier and shaper are therefore designed to provide an
anti-aliasing filter that restricts the frequency content of the
analog signal to frequencies below the frequency of \SI{1}{\MHz}. 
The MicroBooNE anti-aliasing filter is based on a fifth order,
low-pass network designed to obtain an impulse response close to a
Gaussian distribution in the time domain.
The impulse response function in the time
domain is obtained from the inverse Laplace transformation of the
transfer function for the network
\begin{equation}
\label{eq:tran_func}
T(s) = \frac {A_0\cdot C_A} {(p_0 + s)\cdot (p_{i1}^2 + (p_{r1}+s)^2) \cdot (p_{i2}^2 + (p_{r2}+s)^2)},
\end{equation} 
with $s$ being a complex frequency variable.
The parameters in equation~\ref{eq:tran_func} are obtained from a detailed
simulation of the network design and are determined to be:
\begin{equation}
\label{eq:tran_func_par}
\begin{aligned}
p_{r1} &= \frac{1.417}{t_p \cdot C_T}, & p_{r2} &= \frac{1.204}{t_p \cdot C_T}, \\
p_{i1} &= \frac{0.598}{t_p \cdot C_T}, & p_{i2} &= \frac{1.299}{t_p\cdot C_T}, \\
p_0 &= \frac{1.477}{t_p\cdot C_T}, & C_A &= \frac{2.7433}{(t_p \cdot C_T)^4}, \\
C_T &= \frac{1}{1.996}; 
\end{aligned}
\end{equation}
where $A_0$ is the gain parameter and $t_p$ is the 
peaking time constant. $T(s)$ has units of $\frac{\si{\V}}{\si{\C}}$ ${\si{(\Hz)}^{-1}}$.

 

\begin{figure}[!h!tbp]
\includegraphics[width=\figwidth]{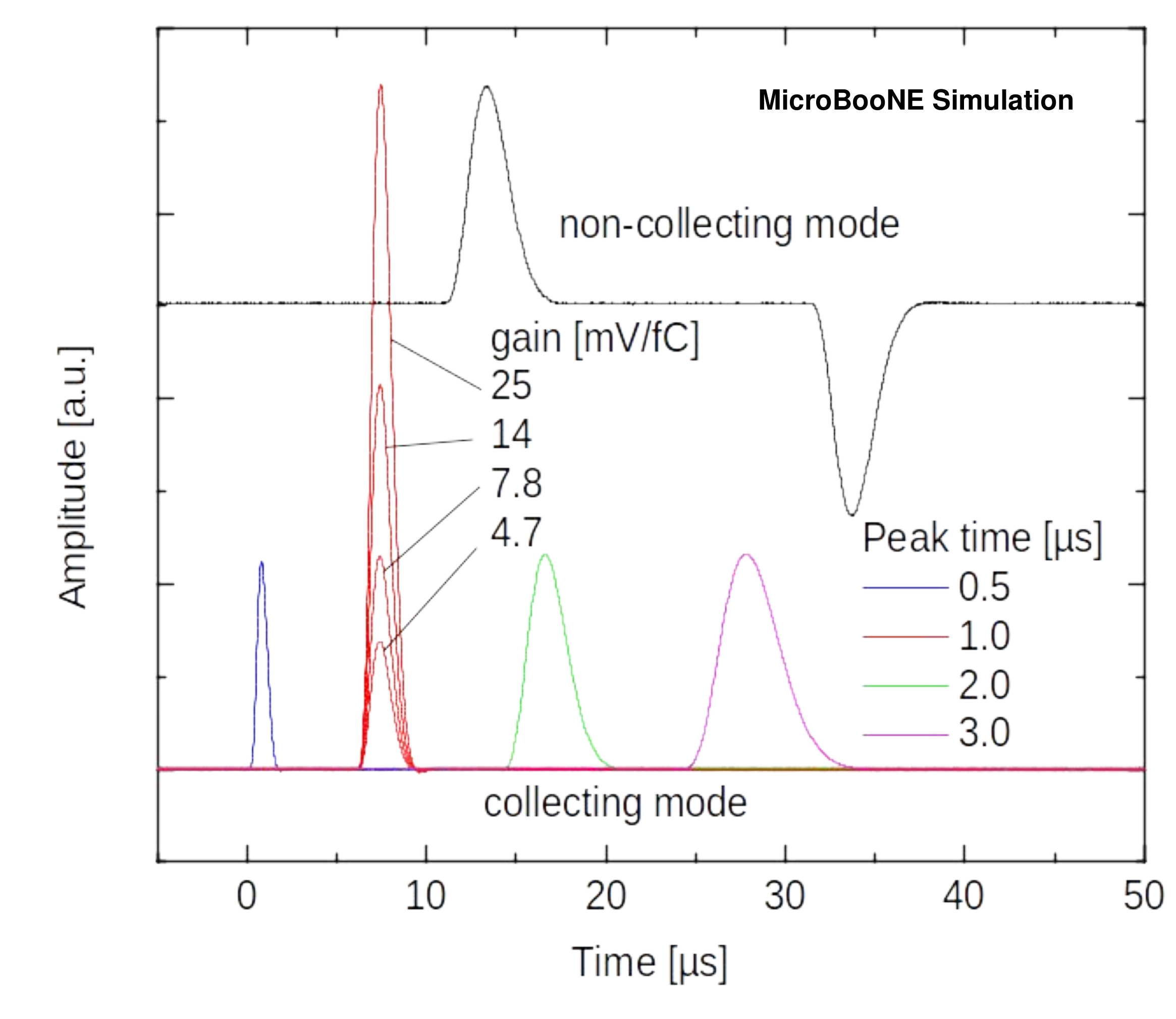}
\caption[resp]{ASIC impulse response functions are shown for four
  peaking times and four gain settings. 
  The example of a calibration pulse (black curve) through injecting a rectangular wave is also shown. 
  Vertical axis units (a.u.) are equivalent to \SI{}{\mV/\fC}.
  The baseline setting is \SI{\approx200}{\mV} for collecting mode and \SI{900}{\mV} for non-collecting (induction) mode.}
\label{fig:resp1}
\end{figure}

The impulse response function in the time domain is shown in
figure~\ref{fig:resp1}.  The front-end cold electronics are 
programmable with four different gain settings 
(\SIlist{4.7;7.8;14;25}{\mV/\fC})
and four peaking time settings 
(\SIlist{0.5;1.0;2.0;3.0}{\us}).
For a fixed gain setting, the peak of the impulse response
is always at the same height independent of the peaking time.
The different gain settings allow for use in applications with differing ranges of input signal strength.
The four peaking time settings are provided to satisfy the Nyquist
criterion at different sampling rates. More details on peaking time
and gain optimization can be found in section~\ref{sec:optimization}.  A
calibration capacitor of \SI{183}{\fF} is installed in the ASIC to allow for
a calibration of the electronic response for every
channel.
Given the overall output dynamic range of \SI{1.6}{\V} (the shape of the
impulse response function is shown in figure~\ref{fig:resp1}) and the
expected level of intrinsic noise, the baselines chosen for the collection plane
and the induction plane (bipolar signals) are \SI{\approx200}{\mV} and \SI{900}{\mV}
respectively. From the test stand and in-situ pulser measurements, it was observed that the channels with a 
baseline of \SI{900}{\mV} have $\approx$3\% lower gain than the channels with a baseline of \SI{200}{\mV}.

\begin{figure}[!h!tbp]
\subfigure[]{
\includegraphics[width=\fighalfwidth]{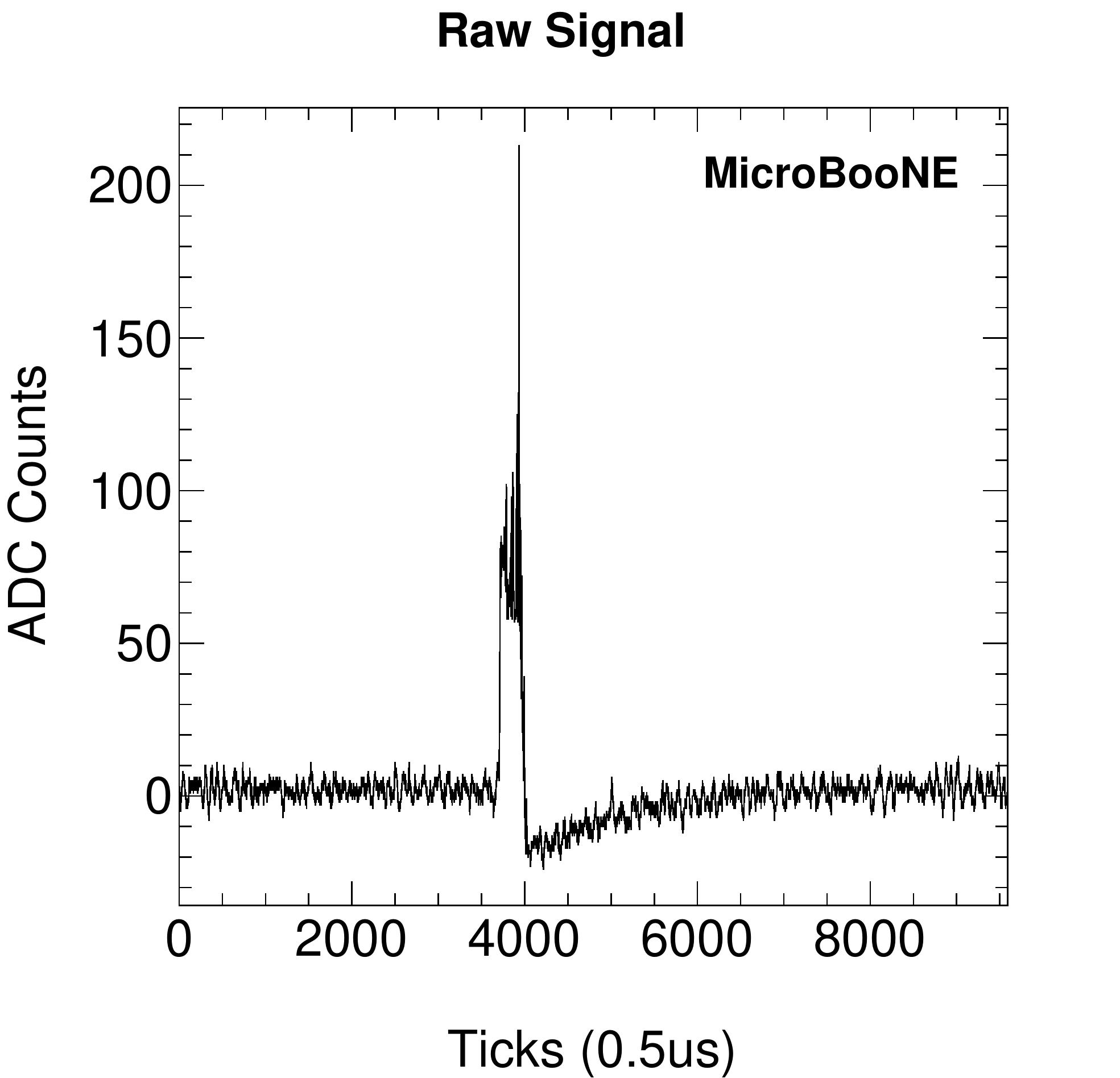}
\label{RC-a}
}
\subfigure[]{
\includegraphics[width=\fighalfwidth]{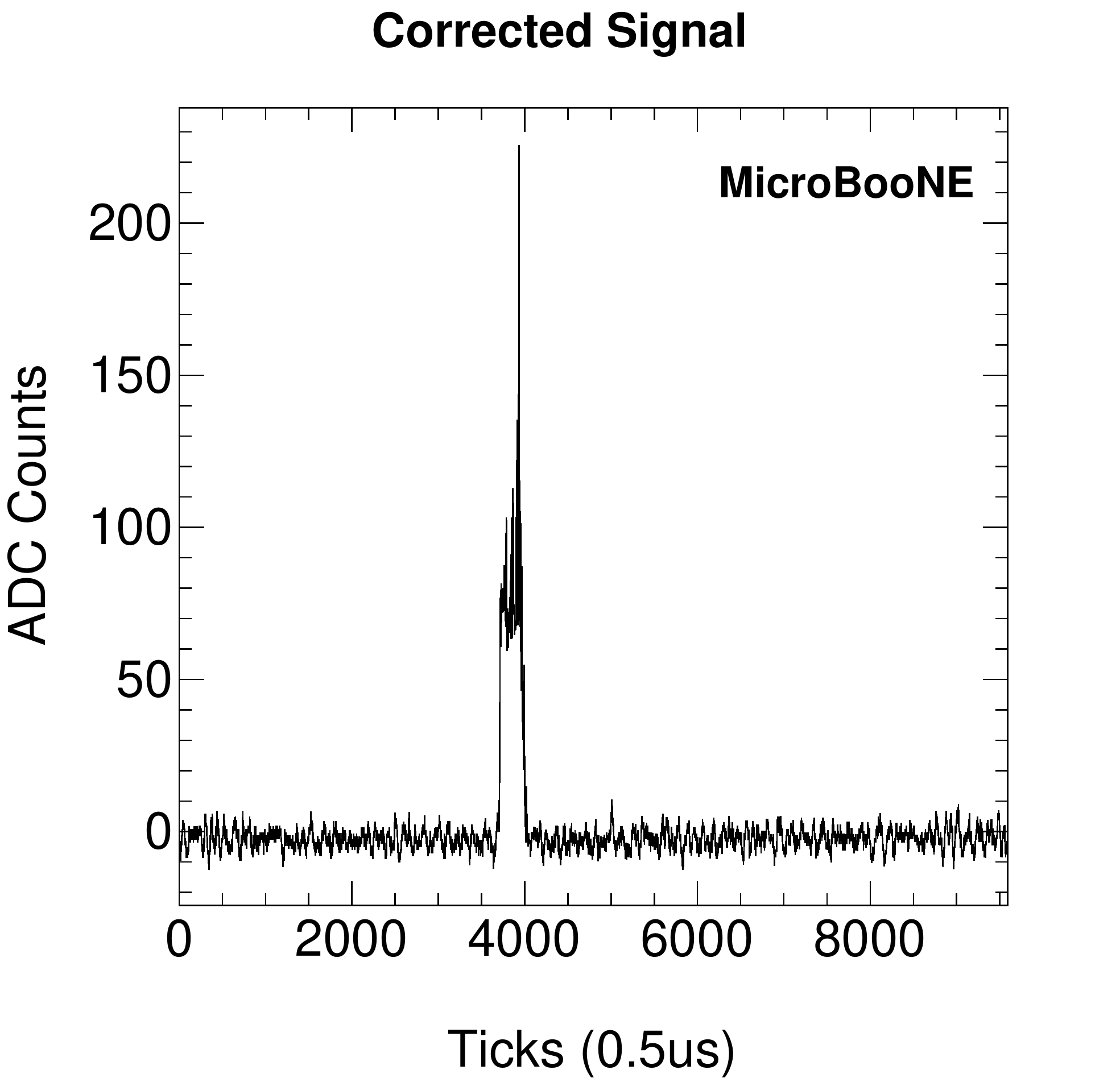}
\label{RC-b}
}
\caption[RC]{Example Y plane raw signal waveforms from data: (a) before correcting, and 
(b) after correcting for the RC shaping effect. This signal 
corresponds to a cosmic muon traveling close to parallel to the collection wires.} 
\end{figure}

For the two RC circuits, the resulting total electronics response is a convolution 
of their individual RC response functions: 
$R_{RC} (t)$ = $\delta(t) - (e^{-t/t_0}/t_0$) with $t_0 = \SI{1}{\ms}$. These RC circuits introduce long tails to the impulse
response that may lead to distortions in reconstructed ionization charge distributions along the drift
direction. Since the time constant is generally long (\SI{1}{\ms}) with
respect to the duration of a typical signal, the effect of this
response function is negligible for most signals. However, in the case
of a large amplitude signal with a long duration (for example, a long,
vertical cosmic muon track that is almost parallel to the
collection wires), the effect of the RC circuits is visible. 
Figures~\ref{RC-a} and~\ref{RC-b} show the signal before and
after correcting for the RC shaping effect through the deconvolution
process~\cite{decon}.  Such a correction is generally effective in
removing this type of distortion.  There are certain
scenarios where such a correction would fail.  For example, if the
signal from a vertical cosmic muon arrived just before the \SI{4.8}{\ms}
readout time, the recorded data would miss the primary signal that is
large and positive, and would only include a long negative tail.
Failing to account for this effect would
lead to a significant baseline distortion for its few occurrences.
Instead, such signals are identified
by examining the very low frequency amplitudes and, where identified, the
slowly varying baseline under these signals is directly removed in the
time domain.

\subsection{ASIC operation optimization}
\label{sec:optimization}

Some flexibility to optimize the ASIC operation in-situ is allowed by
the ability to select among the four gain and four peaking time
settings provided by the cold ASIC.  There are two considerations when
determining the optimal gain setting.  The gain must be chosen low
enough to ensure a large signal, such as from many highly ionizing
protons exiting a neutrino interaction vertex over a small volume,
does not saturate the pre-amplifier.
On the other hand, the gain must be large enough to allow small
signals, such as from MIP track after undergoing
both electron diffusion and absorption across a large drift, to be
well above the gain-independent noise from sources beyond the
pre-amplifier (e.g., noise associated with the intermediate amplifier).
The nominal gain setting for MicroBooNE is \SI{14}{\mV/\fC}.
The saturation limit is more than 45 times larger than the signal due to a MIP 
traveling parallel to the wire plane and perpendicular to the wire
orientation, and more than seven times larger than the signal due to a stopping proton.  
This estimation assumes negligible charge attenuation during the drift.
With this setting, the electronics noise from the intermediate
line-driver amplifier and the ADC is expected to be negligible
compared to the inherent noise associated with the cold ASICs.

Criteria for selecting an optimum peaking time, and ADC sampling rate
are related, as the choice determines the amount of resulting
noise.  The peaking time determines the level of detail of the
original input signal that is retained while the sampling rate
determines how accurately that remaining detail is recorded.
This means that, for best waveform reproduction accuracy, the optimum
choice for the peaking time should scale inversely with the optimum
choice for the sampling rate.
As will be discussed in section~\ref{sec:cold_noise}, the peaking time also
impacts the inherent noise from the pre-amplifier and shaper. An
optimization of the peaking time and the sampling rate therefore
affects both the accuracy with which the signal can be measured, and
the signal-to-noise performance of the readout electronics.


Given the continuous nature of a realizable low-pass filter there
is a necessary trade-off between the amount of bandwidth the filter passes and
the amount of aliasing it allows.  Either the filter attenuates frequencies
below the Nyquist frequency, or it allows some power above the
Nyquist frequency leading to some aliasing.
For MicroBooNE, we chose to avoid aliasing and \textit{oversample}
at a rate higher than strictly required to cover the bandwidth of the
filter.  Oversampling is defined as
\begin{equation}
M = \frac {f_N}{f_{NR}} = \frac {f_s} {4f_{3db}},
\label{eq:oversampling}
\end{equation}
where $f_N$ is the Nyquist frequency and $f_{NR}$ is the frequency 
beyond which the power of signals is negligible after applying the anti-aliasing filter.
$f_s$ is the sampling rate with $f_N$=${f_s}/{2}$ and 
$f_{3db}$ is the frequency where the response of
the anti-aliasing filter is reduced by \SI{3}{\dB}. The cut-off frequency of the signal 
$f_{NR}$ is then estimated to be 2$f_{3db}$.


In MicroBooNE, the sampling frequency is \SI{2}{\MHz} and the shortest
peaking time for the anti-aliasing filter sufficient to provide high
accuracy charge measurement is therefore $t_p = \SI{1}{\us}$. For the
filter response described by equation~\ref{eq:tran_func},
${f_{3db}}\approx\SI{0.25}{\MHz}$, resulting in oversampling of $M =
2$. In principle, samples taken after longer anti-aliasing filters
(peaking time \SI{\approx2}{\us}, \SI{3}{\us} and higher) contain
progressively less and less information about the finer features of
the induced current waveforms. The optimum choice of the sampling
frequency and corresponding peaking time is ultimately determined by
the time scale of the induced currents, i.e., by the electron drift
velocity, wire plane spacing, and diffusion. In MicroBooNE, the
peaking time was chosen to be \SI{2}{\us} instead of \SI{1}{\us} due to the
chosen drift field.  The lower drift field leads to slower electron
drift which has two effects on the inherent time spread of
signals.  First, it leads to a longer feature size of the signal due
to a longer drift near the wire planes.  Second, drifts require more
time on average, allowing for more diffusion which again leads to smoother
signal features.  All together, little information is lost with a \SI{2}{\us}
peaking time, and this has the advantage of a slightly lower inherent
noise than the \SI{1}{\us} setting as discussed in the following section.

\subsection{Front-end ASIC inherent noise}
\label{sec:cold_noise}

In this section, we describe the inherent noise associated with the
front-end ASIC.  The major components of this irreducible noise are:
\begin{itemize}
   \item Series noise associated with the gain mechanism in the 
     first transistor of the charge pre-amplifier. It consists of two components:
     \begin{itemize}
     \item a ``white series''
       noise due to thermal fluctuations in the input transistor, 
     \item a  ``$1/f$ series'' noise (sometimes referred to as ``pink'' noise) due to 
       charge trapping and de-trapping in the input transistor. The $1/f$
       noise dominates at low frequencies.
\end{itemize}
       
  \item ``white parallel'' noise due to the transistor bias current
and resistors providing wire bias voltage.

  \item ``$f$ parallel'' noise which arises from thermal 
  fluctuations in the dielectric components such as the circuit boards and wire carrier boards. 
  Upon integration on the input capacitance this noise acts as a series $1/f$ noise~\cite{vr_signal}. By design,
  this noise is much lower than other sources. 
 
\end{itemize}

The level of noise from the entire electronic readout
chain is expressed as the equivalent noise charge
(ENC) measured in units of number of electrons.
The ENC is defined as the
number of instantaneously collected electrons required so that their peak ADC count
is equal to the root mean square (RMS) of the noise measured, also in the units of ADC counts.
The ENC from the noise sources listed above can be approximated as~\cite{encsim,vr_signal}:
%
\begin{equation}
{\rm{ENC}}^2 \approx {\frac{1}{2} {A_1} \frac{{e^2_n}{C^2_{in}}} {t_p}} +  {A_2 \pi C_{in}^2 A_f} + 
{A_3 \left(q_e I_o + \frac {2 k_B T} {R_b}\right) t_p},
\label{eq:ENC}
\end{equation}

The first two terms represent the white and $1/f$ series
input transistor noise.  Here, $C_{in}$ is the total capacitance at the
input of the ASIC, which includes that of the sense wire, the leads, their
connections to the ASIC and the input transistor gate capacitance. 
Here, $e_n$ is the white series noise spectral density in the unit of \si{\V/\Hz}. 
The series $1/f$ noise spectral density in $\si{\V\per\sqrt{\Hz}}$
is given by $\sqrt{A_f/f}$. Both $e_n$ and $A_f$ are
determined from the ASIC's input  p-channel metal-oxide
semiconductor (pMOS) transistor measurements and depend on
the transistor technology and transistor design (optimization of
transistor electrode dimensions and operating conditions as described
in reference~\cite{veljko}). The two components in the third term represent white
parallel noise, arising from the ASIC bias current, $I_o$ (shot noise)
and the wire bias resistor (thermal noise), where $k_B$ is the
Boltzmann constant, $q_e$ is electron charge, $T$ is temperature and
$R_b$ is resistance of the wire bias resistor\fixme{MB: dont we have exact numbers
  for $R_b$ - can that be provided along with approximate values for
  $A_f$ and $e_n$ from the reference?}.  The $t_p$ in equation~\ref{eq:ENC}
is the peaking time of the anti-aliasing filter. The dimensionless
form coefficients for the semi-Gaussian impulse response functions,
such as given by equation~\ref{eq:tran_func} and~\ref{eq:tran_func_par},
are $A_1$=2.0, $A_2$=0.9 and $A_3$=1.0~\cite{vr_signal}.

\begin{figure}[!h!tbp]
\center
\includegraphics[width=\figwidth]{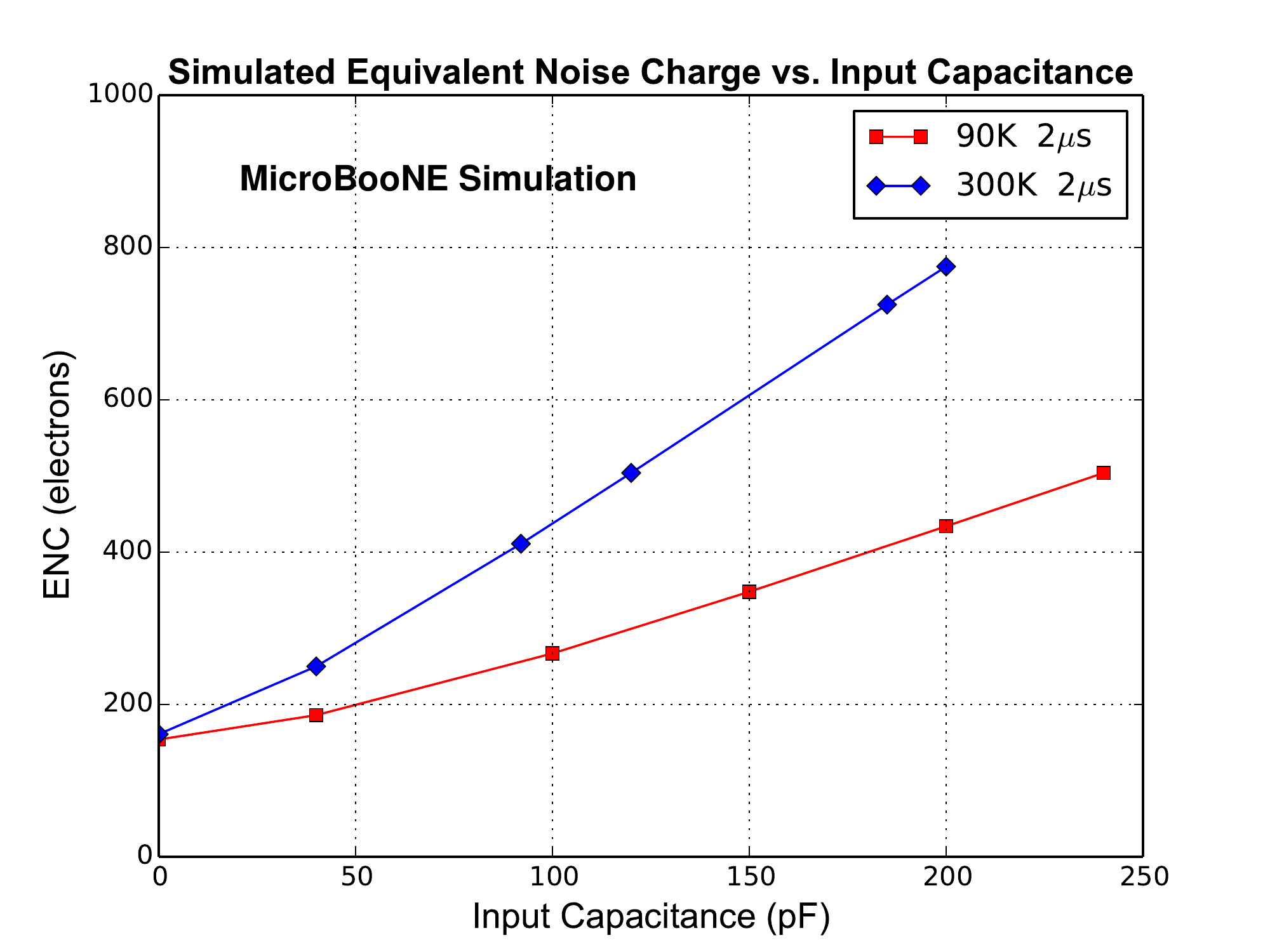}
\caption[ENCvsCin]{Simulated equivalent noise charge vs. input capacitance for $T = \SI{90}{\kelvin}$ (red) and $T = \SI{300}{\kelvin}$ (blue).}
\label{fig:enc_cin}
\end{figure}

At temperatures of \SIrange{77}{89}{\kelvin}, 
charge carrier mobility in silicon
increases and thermal fluctuations decrease with $k_B T / q_e$.
This results in higher gain, higher speed, and lower noise. 
The series $1/f$ noise and CMOS transistor white
series noise amplitude, $e_n$ at \SI{77}{\kelvin} is a factor of 2
lower than at \SI{300}{\kelvin}.
%
%
Therefore, operation of the front-end
ASIC at LAr temperatures significantly reduces the inherent noise.  The ASIC has been
designed to minimize the white parallel noise by limiting
the bias current; the lower limit has to be set so as to avoid
pre-amplifier saturation by the currents induced by sense wire
motion (section~\ref{sec:chirping}).  Therefore, the ENC
due to the first transistor noise is the dominant noise source in the
cold ASIC~\cite{Radeka:2011zz,veljko}. For peaking times $>\SI{1}{\us}$, 
the ASIC noise at \SI{77}{\kelvin} is dominated by $1/f$ noise from
the first transistor. For a capacitance of \SI{220}{\pF} connected to the
input, and the peaking time of \SI{2}{\us},
the total ASIC ENC measured is $\approx$550 electrons at
\SI{77}{\kelvin}~\cite{asic}.  Out of this, only $\approx$150 ENC is from white parallel
noise. These results are in good agreement with the expectation from the
simulation as shown in figure~\ref{fig:enc_cin}.

The capacitance of the sense wires contribute substantially to the
input capacitance $C_{in}$ and dominates in the case of the longest
wires.
Two estimates of wire
capacitance are given here.  The first follows
reference~\cite{Erskine:1972tt} in the case of a parallel wire grid,
symmetrically placed between two conducting planes. The capacitance per length $C_{wire}$ is given by:
\begin{equation}\label{eqn:cin}
  C_{wire} = \frac{2\pi \epsilon_o \epsilon_r}{(\pi G/W)-\ln(\pi d/W)},
\end{equation}
where $\epsilon_0$ is the permittivity of free space, $\epsilon_r$
is the relative dielectric constant for liquid argon, $W$ is the wire spacing, $G$ is
the plane gap, and $d$ is the wire diameter. For MicroBooNE's wire spacing and 
\SI{150}{\micro\meter} 
wire diameter, 
this corresponds to a capacitance per unit wire length of \SI{\approx17}{\pF/\meter}. 
This number is expected to be smaller
than its actual value since equation~\ref{eqn:cin} assumes
an electrode geometry of parallel wires between two parallel plates, 
instead of the actual multiple wire plane geometry. 
The second estimate is based on a 2-D finite-element analysis~\cite{VanEsch:2006bx}
and yields a capacitance value of \SI{\approx20}{\pF/\meter}
for U and Y wire plane and \SI{21}{\pF/\meter} for the middle V wire plane.

From equation~\ref{eq:ENC}, a general formula characterizing the observed
front-end readout noise level as a function of wire length in terms of
RMS ADC counts can be written as:
\begin{equation}
\begin{aligned}
  \textrm{ENC} \propto \textrm{Noise}_\textrm{RMS}~\textrm(\rm ADC) = \sqrt{x^2 + {(y  + z \times L)^2}},
  \label{eqn:rms_fit1}
\end{aligned}  
\end{equation}
where $x$ corresponds to the third term in equation~\ref{eq:ENC} and is due to the white
parallel noise, and noise from front-end readout components other than the
ASIC, such as the noise associated with the intermediate amplifier and
the ADC. The second term is due to the series noise, with $y$ arising
from noise due to the transistor gate capacitance and capacitance of
the wire-to-ASIC connections. The last term is due to the wire
capacitance, which is proportional to the wire length $L$. All terms
are expressed in ADC units.  As discussed in the previous sections,
the noise from components other than the cold ASICs in the front-end
readout chain is inversely proportional to the overall gain prior to
the ADC when converted into ENC.

Finally, the inherent ADC noise, including quantization noise, is such
that it provides an equivalent number of bits of 11.3.  This
corresponds to a noise level of $\sigma = 1.6 \times (1/\sqrt{12}) \approx$ 0.46 counts of the 12 bit ADC.

In the following sections, we describe the status of TPC readout, 
characteristics of the excess noise seen in MicroBooNE TPC, and a 
comparison with the expected inherent noise.

\section{Status of the TPC readout post-installation}
\label{sec:non-func}

A study of the status of the TPC readout channels in MicroBooNE was
conducted after installation and filling of the cryostat with liquid argon. The
pedestal distribution during regular data taking and the ASIC impulse
response using calibration data taken with the external
pulser are used to identify both operational and non-functioning
channels. It was discovered that $\approx$10\% of the channels were
non-functional.\footnote{For event reconstruction using two wire planes, 
the resultant inefficiency is estimated to be $\approx$3\%.} 
In the following sections, the causes of various
channel failures are detailed. The operational status of the
channels are taken into account in the further steps in data analysis. 

\subsection{Misconfigured channels}
\label{sec:misconfig}

The \num{8256} channels reading out the MicroBooNE TPC are grouped
into 516~ASICs~\cite{uboone_1} (16~channels per ASIC).  As discussed
in section~\ref{sec:optimization}, the gain and peaking time settings for
the ASICs during production data taking were selected to be
\SI{14}{\mV/\fC} and \SI{2}{\us}, respectively. By analyzing raw
data in the time and frequency domain and using the pulser calibration
signals to determine the gain, it was found that 14~ASICs (224
channels) were instead consistent with the factory settings of a gain
of \SI{4.7}{\mV/\fC} and a peaking time of \SI{1}{\us}.  All of these
channels read out the first induction (U) plane.  The cause of this
mis-configuration is believed to be due to damage to the configuration
signal lines on these ASICs from electrostatic discharge (ESD) during
detector installation.

\begin{figure}[!h!tbp]
\center
\includegraphics[width=\figwidth]{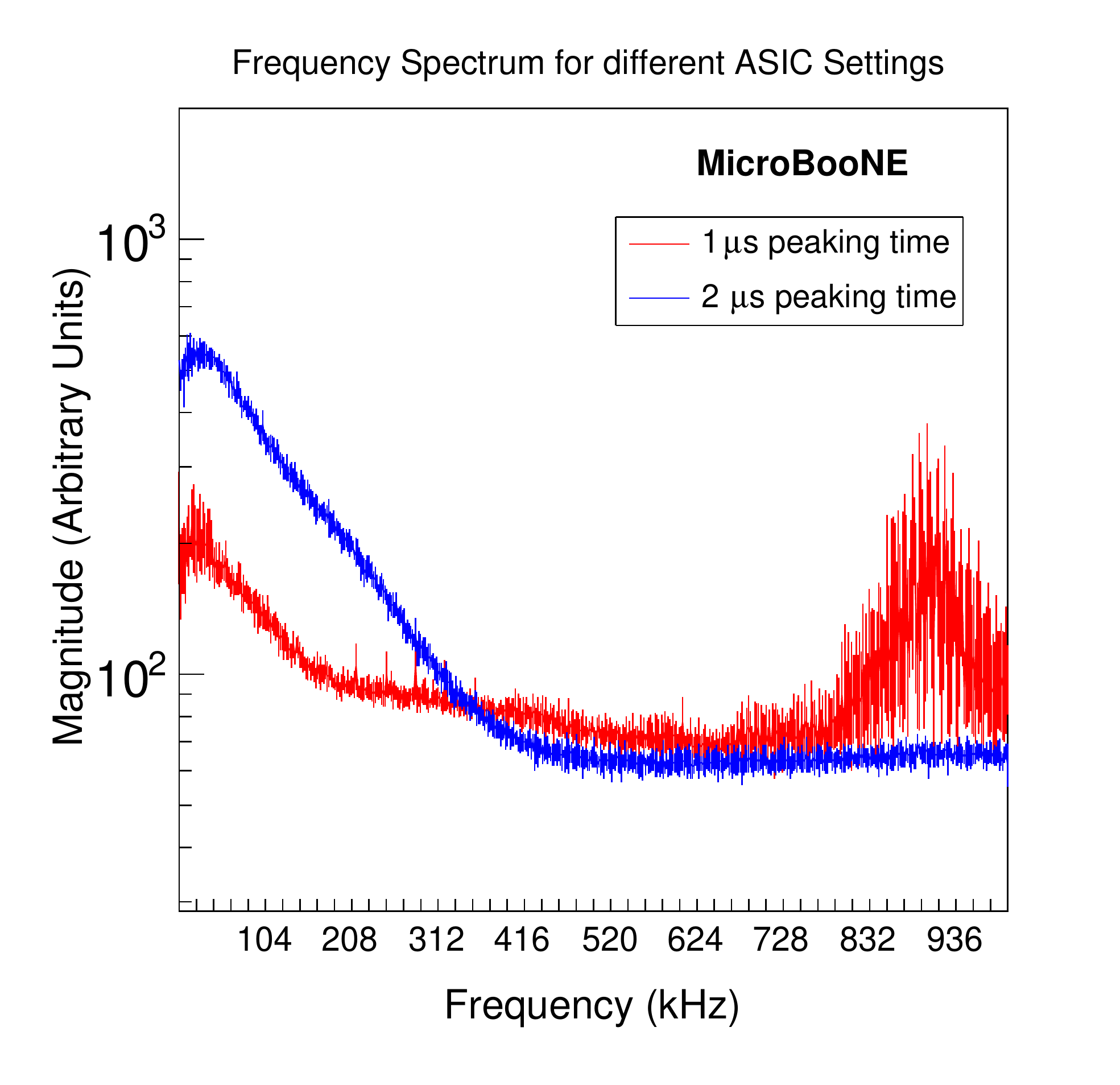}
\caption[mis-con]{The average frequency domain magnitude for channels
  with \SI{1}{\us} peaking time and \SI{4.7}{\mV/\fC} gain (red) and with \SI{2}{\us} peaking time and
  \SI{14}{\mV/\fC} gain (blue). The magnitude represents the amplitudes of
  various frequency components. The average is taken over $\approx$200 U
  plane channels and has smoothed out random factors associated with
  the noise spectrum for individual channels. The peak 
  \SI{\approx900}{\kHz} for the \SI{1}{\us} setting is due to the high frequency
  burst noise described in section~\ref{sec:zig-zag}.}
\label{mis-con-2}
\end{figure}

Since the expected electronic noise characteristics are different for
different peaking times, these misconfigured channels can be
identified in the frequency domain. As shown in figure~\ref{mis-con-2},
the average frequency content of signals recorded by channels with a
\SI{2}{\us} peaking time is clearly different from that of channels
with a \SI{1}{\us} peaking time. The higher peaking time acts as a
low pass filter, removing the highest frequency components.
The relative difference in the heights of the two low frequency peaks
are evident and consistent with the two gain and peaking time settings
as well as the noise from non-ASIC sources (e.g., the ADC and line-driver amplifier).

Although these channels are misconfigured, they are usable because a signal
can still be discerned from the electronic noise.  The shaping
produced by the misconfigured settings (\SI{4.7}{\mV/\fC} and
\SI{1}{\us}) is corrected by the shaping corresponding to the intended
settings (\SI{14}{\mV/\fC} and \SI{2}{\us}).  This correction is done
through an offline noise filter that allows the waveforms from these
channels to be treated in the same manner as those from other channels
by the procedures that are applied later in the analysis chain.

The misconfigured channels are associated with larger noise than the
other channels. First, the \SI{1}{\us} peaking time setting leads to
\SI{\approx10}{\percent} higher inherent ASIC noise compared to the
\SI{2}{\us} peaking time~\cite{asic} (see section~\ref{sec:RMS}).  More
importantly, as discussed in section~\ref{sec:optimization}, the lowest
gain setting leads to a much larger relative contribution of the electronics
noise originating downstream of the cold ASIC from the
intermediate amplifier and the ADC.

\subsection{Shorted channels}
\label{sec:dead}

At the start of the commissioning, all wire bias voltage lines were
tested and exhibited the expected high resistance to ground (in the
\si{\GOhm} range). The sense wire bias voltage is supplied to the
wires through the signal feedthroughs~\cite{uboone_2}. Upon
application of the bias voltage, three signal feedthroughs started
drawing more current than the voltage source could supply. In
coincidence, the noise increased on a number of U channels 
by a factor
of $\approx$20 and the external pulser signal was reduced by $\approx$30\%.
An anomalous response was also observed on a single V plane channel.

These observations could be explained by a single V
wire coming in contact with a number of U wires, thus establishing a DC
electrical contact across different feedthroughs. The diagnosis by DC
measurements was facilitated by the wire bias distribution
arrangement.  Each U or Y wire receives its bias voltage via a
\SI{22}{\MOhm} resistor.  On the other hand, each V wire has zero bias
as it is connected directly to the protection diodes at the input to
the ASIC. Thus, a measurement of the current vs. bias voltage at the
feedthrough indicates how many \SI{22}{\MOhm} resistors appear
connected in parallel. This number is interpreted to be the number of
U~wires from a particular feedthrough that are in contact with the
V~wire.  The derived number of shorted wires during production data taking 
is consistent with the 287 non-functioning
U~channels, of which 259 channels have no signal and 28 channels have a high level of noise.

These U~wires are not expected to retain their bias voltage and thus
disturb the nearby electric field. As a result, some amount of 
drifting ionization electrons that would otherwise pass by these
U~wires is instead expected to be collected on them.  This leads to a
smaller signal amplitude in the subsequent nearby V and Y~wires.  This
behavior is confirmed from analysis of the TPC signals from these
channels. The MicroBooNE detector is welded shut making it 
impossible to easily access the interior of the cryostat for any diagnostics. 
A visual scan on the TPC wire plane was immediately performed~\cite{touching}. 
However, as no direct contact between the V and U wires was observed, 
the exact location of the short is not known at the moment of writing of this paper.

Beside these U wires, there is a group of Y wires that are identified to be shorted 
to ground through the leakage current for a bias voltage power supply. 
As a result, the ionization electrons that would otherwise be
collected by these Y~wires are instead collected by the V~wires
crossing in front of them.  This was confirmed by observing unipolar
signals in these V~wires instead of the expected bipolar signal.
Signals from ionization electrons passing by V~wires at
locations distant to the grounded Y~wires do produce the expected
bipolar induction signals.  This indicates that the problems arising from the
grounding of the Y~wires are local to these wires and do not extend to
the entire length of the crossing V~wires. 
The exact cause of this short is also not known at the writing of this paper.

\subsection{Periodic saturation of ASICs}
\label{sec:chirping}

During the initial commissioning period of the MicroBooNE TPC, 
$\approx$10--15\% of the induction planes channels were observed to
display a periodic ``saturation'' behavior whereby the channel would
be essentially non-functional for a period of time and then recover.

\begin{figure}[!h!tbp]
\center
\subfigure[]{
\includegraphics[width=\fighalfwidth]{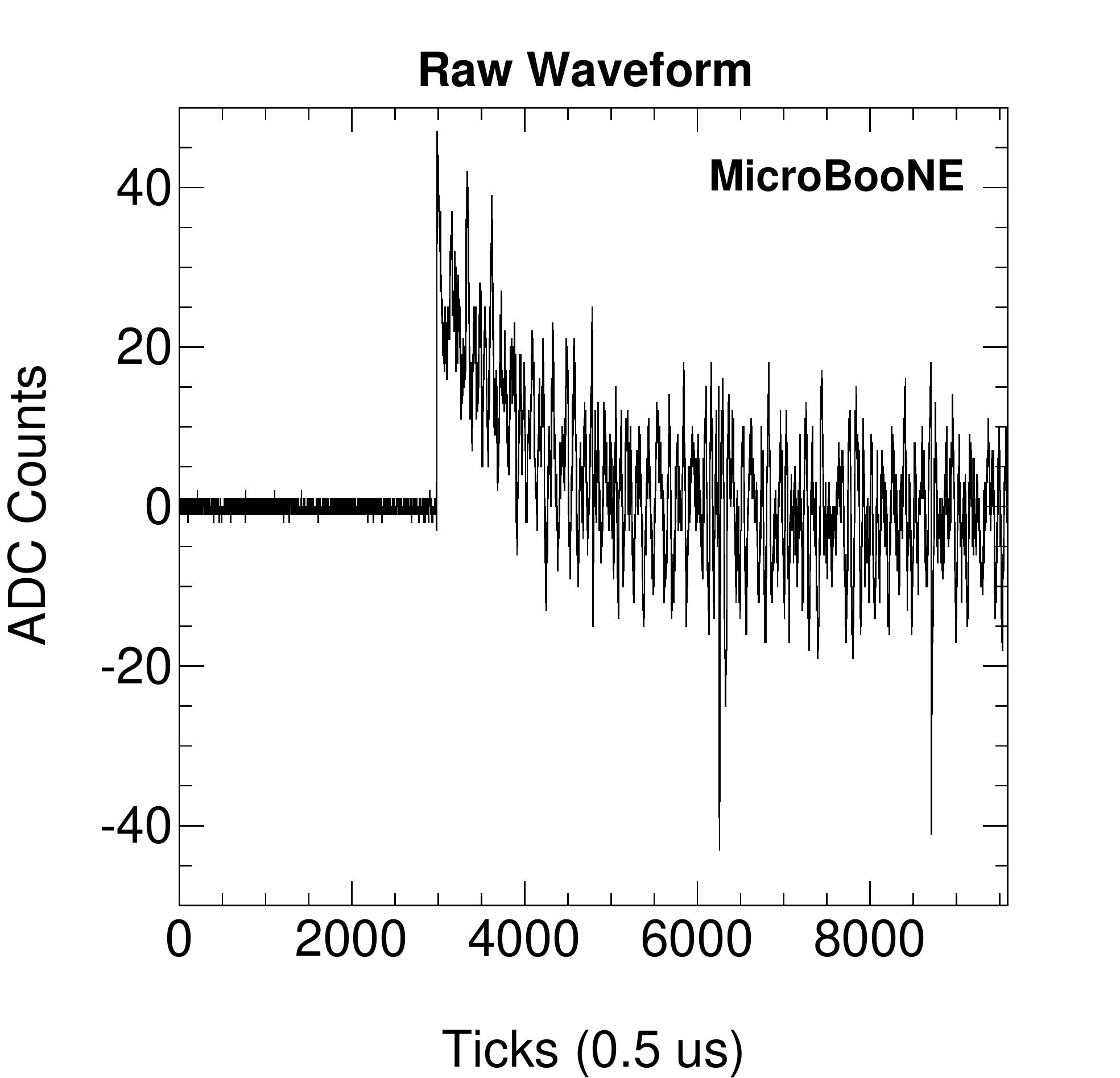}
\label{chirping-a}
}
\subfigure[]{
\includegraphics[width=\fighalfwidth]{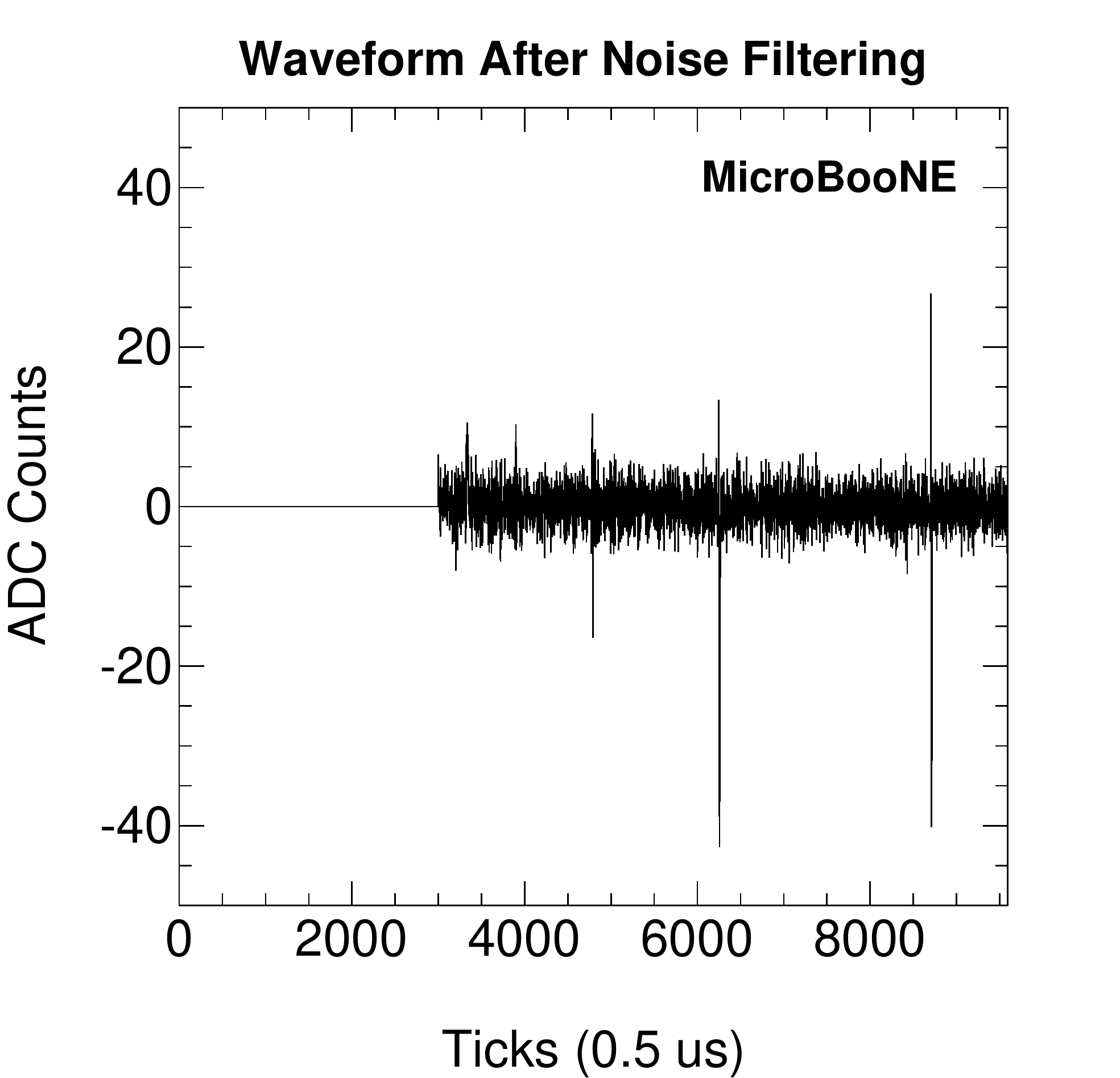}
\label{chirping-b}
}
\caption[chirp_1]{An example V plane raw waveform: (a) after the recovery
  of an ASIC from saturation, and (b) the same waveform after the noise
  filter. In (a), the region corresponding to the ASIC saturation
  (low RMS region between 0 - 3000 Ticks) is clearly seen. In (b), the distortion of the baseline is removed 
  by the noise filtering.}
\end{figure}

Figure~\ref{chirping-a} shows a waveform from an ASIC channel that has
become saturated for part of the readout window. The saturation behavior is characterized by very low gain, low noise and no response to the external pulser. Some time after
saturation the channel recovers to a normal state. Just after
making a gross recovery, there is a period when the baseline is
distorted due to the RC circuits as described in section~\ref{sec:RC}.
Saturation occurs sporadically and its incidence is higher when the
wire bias voltage is higher. No saturation is seen for wires at zero
bias voltage. Direct oscilloscope measurements at the output of the
intermediate amplifier of several U and V channels that go into
saturation indicate they do so with a distribution of average
frequencies less than \SI{15}{\Hz}.

ASIC saturation occurs when a positive input current exceeds the
bias current that keeps the ASIC input circuit in equilibrium.  With
a large positive current, the preamplifier is driven against its lower
limit (near zero) and the signal gain becomes very low.  This was
confirmed by a dedicated bench test. The ASIC bias current should be
maintained as low as possible as it contributes to the white parallel
noise (shot noise) as discussed in section~\ref{sec:cold_noise}. The most
likely cause of a large external current arises from the wire motion in the presence of
an electric field.

A detailed study showed that, at low ASIC bias current settings, even a
small wire displacement is sufficient to saturate the ASIC.  For
example, a V~wire displacement of \SI{\approx100}{\micro\meter} (less
than the \SI{150}{\micro\meter} wire diameter) at frequencies in the
range of \SIrange{1}{20}{\Hz} can induce a charge greater than
\SI{50}{\pC} to saturate the ASIC that is biased at \SI{100}{\pA}.

One source of this saturation that has been considered and ruled out
is a microphonic effect (noise due to mechanical vibrations). This
source is unlikely cause as the damping provided by the liquid
argon would not allow for any resonant excitation. Instead, more
likely cause is convection of the liquid argon. Saturation is less
likely on shorter wires because wire displacement is proportional to $l^n$,
where $l$ is wire length and $n=2, 3$ depending on the displacement
force distribution. This is consistent with the observed behavior.

The cold ASIC has two bias current settings (\SI{100}{\pA} or
\SI{500}{\pA}). As shown in equation~\ref{eq:ENC}, the bias current
contributes to the overall noise level. The \SI{100}{\pA} setting is
intended for short wires in induction wire planes as it gives a
smaller shot noise. The \SI{500}{\pA} setting is the default setting
for the majority of wires, which gives \num{\approx80} and \num{110} ENC 
at \SI{1}{\us} and \SI{2}{\us} peaking time,
respectively. During the initial MicroBooNE data taking, the ASIC bias
current was inadvertently set to \SI{100}{\pA}. It was observed
that a large fraction of channels on the induction planes (10-15\%)
and a few channels on the collection plane experienced quasi-periodic
saturation. After resetting the ASIC bias current to \SI{500}{\pA},
the incidence of saturation was reduced to a few tens of wires.  The
remaining intermittent saturation of the ASICs is time dependent and more
details will be discussed in
section~\ref{sec:time_dep}.

Remaining saturation effects can be corrected offline. The identification of the saturation region in the offline data
processing is based on finding that the local RMS of the pedestal is
less than 1 ADC count within 20 time ticks.  The regions of
individual digitized signals corresponding to the saturation period
are then identified on an event-by-event basis and the baseline distortion
just after recovery is corrected. Figure~\ref{chirping-b} shows the waveform after applying an offline
noise filter and removing the baseline distortion.

\subsection{Summary and time dependence of non-functioning channels}
\label{sec:time_dep}

The number and causes of the non-functioning channels in MicroBooNE are summarized
in table~\ref{tab:bad_noisy}. First, there are $\approx$20 
channels that sporadically suffer from ASIC saturation after
the reset of the bias current to \SI{500}{\pA}.  The exact set of
saturating channels varies with time since the saturation frequency
varies for different channels, and changing detector conditions can change the
liquid flow, which is suspected to cause the saturation.
Second, 6 ASICs are not connected to TPC wires due to 
an installation error.
Third, 19 ASICs cannot be initialized properly inside the liquid argon 
(so-called ``start-up'' problem).  This issue is believed to be
caused by the bandgap reference (BGR) circuit in the ASICs.  A BGR is
a temperature independent voltage reference circuit widely used in
integrated circuits. It produces a fixed constant voltage regardless
of power supply variations, temperature changes, and circuit load.
For the ASICs with this start-up problem, the BGR circuit is suspected
to be stuck in an incorrect initial state when the power supply ramps 
from 0 to \SI{1.8}{\V}.  This is likely due to a larger
transistor current in the cold than in the warm.  Last, there are 
126 Y channels with 116 channels having no signal and 10
channels with high noise located near the shorted wires between the V and
Y plane and 287 U channels with 259 channels having no signal
and 28 channels with high noise located near the shorted wires between the
V and U planes. There are another 36 channels with high noise that have not
been associated with shorted wires.

\begin{table}[!htbp]
  \center
  \caption[Bad channels]{Summary of the non-functioning channels for
    Run 3455 Event 6. A total of $\approx$862 channels are considered
    non-functioning. MB stands for front-end motherboard.}
    \vspace{2mm}
  \label{tab:bad_noisy}
  \begin{tabular}{|c|l|}
    \hline
    \# non-functioning channels & Reason \\ \hline
    $\approx$20 & ASIC saturation\\
    96 & 6 ASICS on one MB not connected to wires\\
    304 & 19 ASICs due to start-up problem\\
    126 & channels sorrounding U-Y shorted wires with 10 noisy channels \\
    287 & channels sorrounding U-V shorted wires with 28 noisy channels \\
    36  & noisy channels not located near the shorted wires \\
    \hline
  \end{tabular}
\end{table}

\begin{figure}[!h!tbp]
\center
\includegraphics[width=\figwidth]{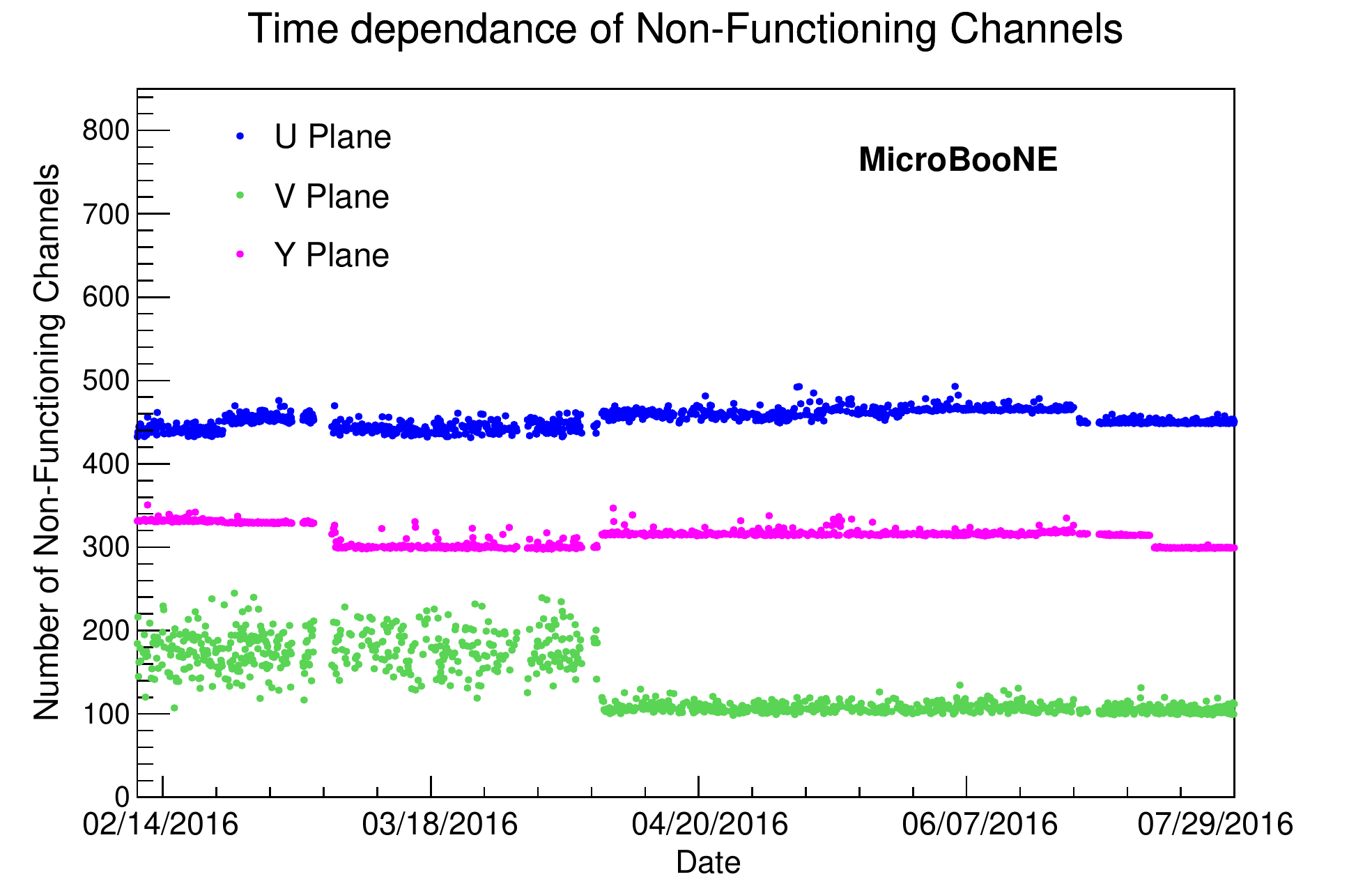}
\caption[dead_time]{Average number of non-functioning channels taken over $\approx$10 events per run as a function of time.}
\label{dead_time}
\end{figure}

Finally, the time dependence of the total number of non-functioning
channels is shown in figure~\ref{dead_time}.
The operational status of the channels was checked over a period
from February to July, 2016.  Each data point in the figure is averaged
over a $\approx$10 event sample from a given run. The number of
non-functioning channels for the U and Y planes is stable except for a period 
at the start of March 2016. For a roughly one month period, the Y plane
had a reduced number of non-functioning channels.  This is due to the
temporary recovery of two ASICs with the ``start-up'' issue that returned to 
a non-functioning state.  Also shown in figure~\ref{dead_time}
is a decrease in non-functioning channels in V plane around the
beginning of April.  This is correlated with a replacement of a faulty
cryogenic pump.  After replacing the cryogenic pump, the occurrence of
ASIC saturation became negligible. This fact suggests that the convection of the liquid argon 
might be the likely cause of saturation as described previously in section~\ref{sec:chirping}.


\section{Identification and filtering of MicroBooNE TPC excess noise}
\label{sec:types}

In section~\ref{sec:cold_noise}, we described the noise inherent to the electronics. 
Excess noise, beyond the expected inherent
noise, is observed in MicroBooNE. Figure~\ref{excess_1} and~\ref{excess_2} shows 
frequency spectra from U plane channels with different gains and
peaking times to illustrate this excess noise. The excess noise is
categorized into three types, ordered in decreasing importance:

\begin{figure}[!h!tbp]
\center
\subfigure[]{
\includegraphics[width=\fighalfwidth]{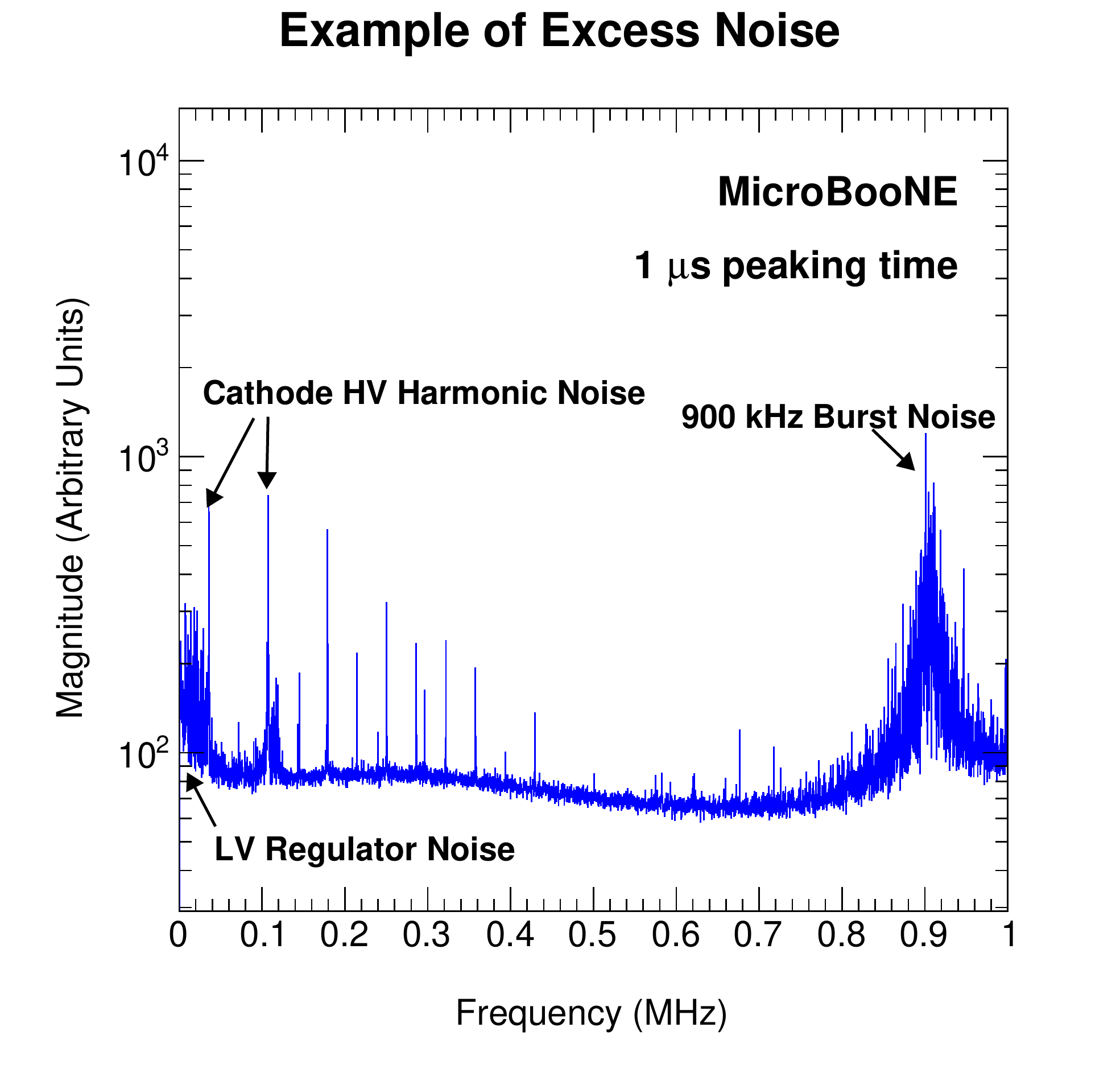}
\label{excess_1}
}
\subfigure[]{
\includegraphics[width=\fighalfwidth]{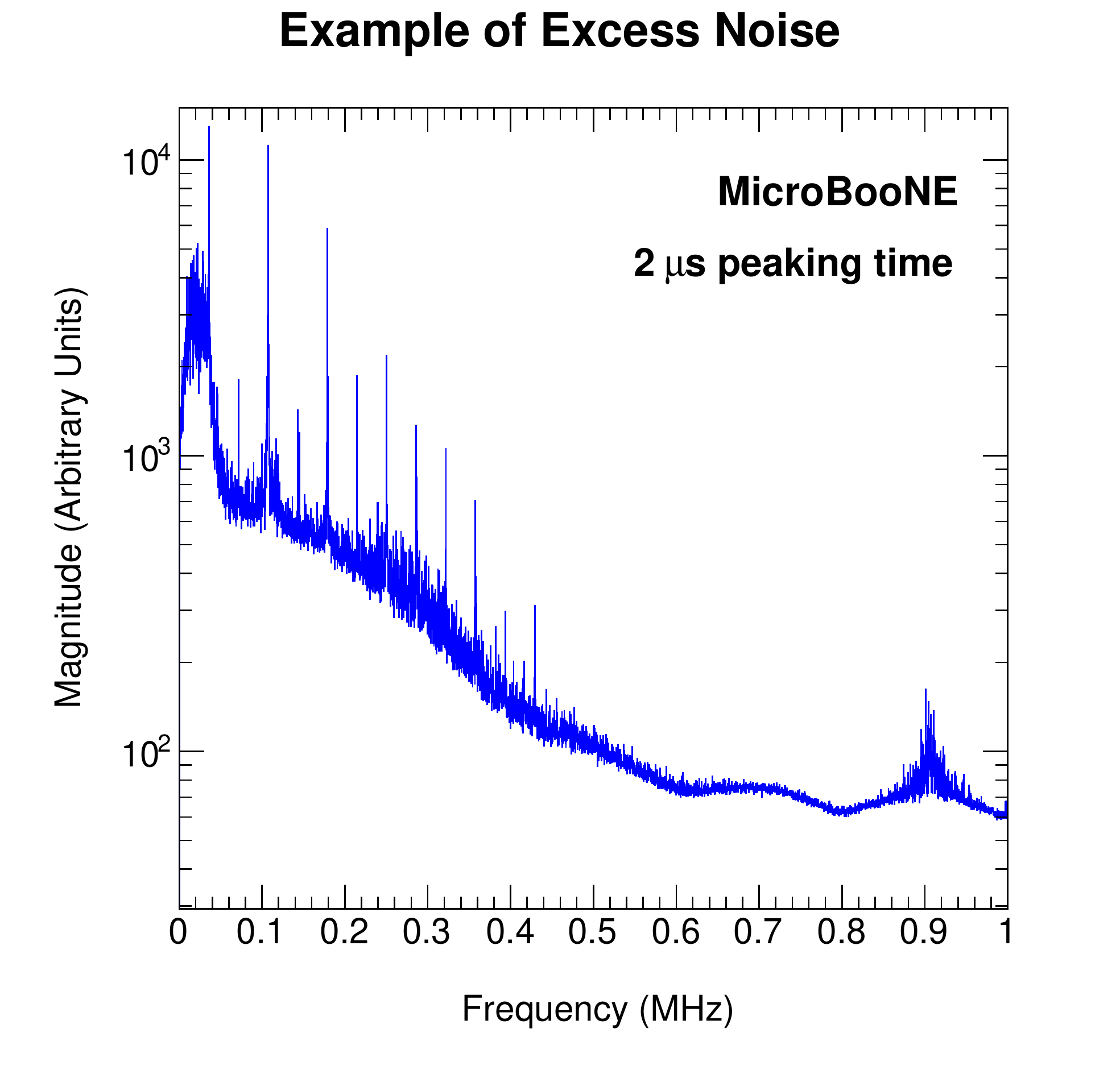}
\label{excess_2}
}
\caption[excess]{Average noise magnitude in the frequency domain: (a) for
  \SI{1}{\us} peaking time and \SI{4.7}{\mV/\fC} gain settings, and (b) for 
  \SI{2}{\us} peaking time and \SI{14}{\mV/\fC} gain settings. 
  Three sources of excess noise are illustrated on the left
  panel: Low voltage (LV) regulator noise, cathode HV harmonic noise
  (\SI{36}{\kHz} and \SI{108}{\kHz} noise), and \SI{900}{\kHz} burst
  noise.}
\end{figure}

\begin{itemize}

\item Noise induced by the low-voltage regulators: In
  section~\ref{sec:coherent}, the noise originating from the low-voltage
  regulators supplying the ASIC operating voltages is described. This
  noise is coherent across all the channels supplied by a regulator
  and shows up at the low end of the frequency spectrum
 ($\lessapprox\SI{30}{\kHz}$).

\item Noise induced by the cathode high voltage power supply: In
  section~\ref{sec:harmonic}, the noise induced on the anode wire plane
  by voltage fluctuations in the cathode potential, such as ripple
  from the HV supply is described. The two 
  highest peaks are near \SI{36}{\kHz} and \SI{108}{\kHz}. 
  The effect of the HV noise is
  largest in the first induction plane (U) where it affects all the
  wires. The noise appears at a reduced level in the second
  induction wire plane (V), and at an even lower level in the collection
  wire plane (Y).

\item \SI{900}{\kHz} burst noise: In section~\ref{sec:zig-zag}, the
  \textit{burst} noise observed at frequencies around \SI{900}{\kHz} is
  described. This type of noise has a clear position dependence and
  has a burst nature.  While its exact source has not yet been confirmed,
  evidence is pointing towards the PMT high voltage supply or the
  interlock system power supply.

\end{itemize}

\noindent In the following subsections, these three types of excess
noise are described in detail and compared to the
inherent noise in order to illustrate their impact on TPC
performance. An offline filter has been developed to remove
most of the excess noise. Its performance on each type of noise is described below. 
The filter has a negligible impact on the signal
as described in the final part of this section.

\subsection{Low frequency noise from voltage regulator}
\label{sec:coherent}

The most significant excess noise observed in MicroBooNE is the low
frequency noise that is injected via the low voltage power input to
the cold ASIC. More specifically, it is due to the voltage regulator
used to provide a stable voltage for the cold ASIC. This noise source is most pronounced
at low frequencies ($\lessapprox\SI{30}{\kHz}$) as shown in
figure~\ref{excess_1}. While the regulator noise is very
low, it still exceeds the noise of the pMOS input transistor.  The
source electrode of this transistor is driven by the regulator voltage
with respect to the gate electrode that is connected to a sense
wire. The resulting injected noise is thus proportional to the wire
capacitance. This noise is typically found to be correlated across 192
channels on the same service board,\footnote{The service board mounted on the top of each signal feedthrough, provides regulated low voltage, control, and monitoring signal to the analog ASICs. It also provides pulse injection to the preamplifiers for precision calibration.}
that
contains the same low voltage regulator. 
Usually,
the correlation is highest within the 48 channels covering three ASICs.

To mitigate this coherent noise, a \textit{correction waveform} is
constructed on a per sample basis and across sets of 48 contiguous
channels. Initially, each sample of this waveform is set to the
median of the corresponding sample tick across the 48 channels.
Nominally, this correction waveform would then be subtracted from each
of the 48 channels. However, certain distributions of ionization electrons,
specifically those from tracks parallel to the wire planes,
would otherwise be suppressed by this simple subtraction. To protect these signals, 
the signal height and shape of the correction waveform are examined to determine if they
contain any regions above expected noise level.\footnote{The signal shape was taken into account through a deconvolution technique using a simulated signal shape for parallel tracks and fast fourier transformation. The threshold of signal height and deconvoluted signal height are all five times the corresponding noise RMS.}  Then, two pad regions extending 7 to 10
ticks on either side of this initial region (20 ticks towards early time for the U plane) are included. 
The central
region of the correction waveform is then zeroed and the pad regions
are linearly interpolated to zero.  Finally, the resulting correction
waveform is subtracted, tick-by-tick, from each of the original 48 channels.

For certain channels, the correction waveform requires additional
construction before it is applied. Some of the channels on the edges
of a service board recorded a larger amount of coherent noise. To
reduce this, a scaling adjustment of the correction waveform is
performed during subtraction for the channels on the edges of a service board. 
The scaling factor is calculated to be
the ratio of the correlation coefficient of a single channel and the
average correlation coefficient of the 48 channels.

\begin{figure}[!h!tbp]
\center
\subfigure[]{
\includegraphics[width=\fighalfwidth]{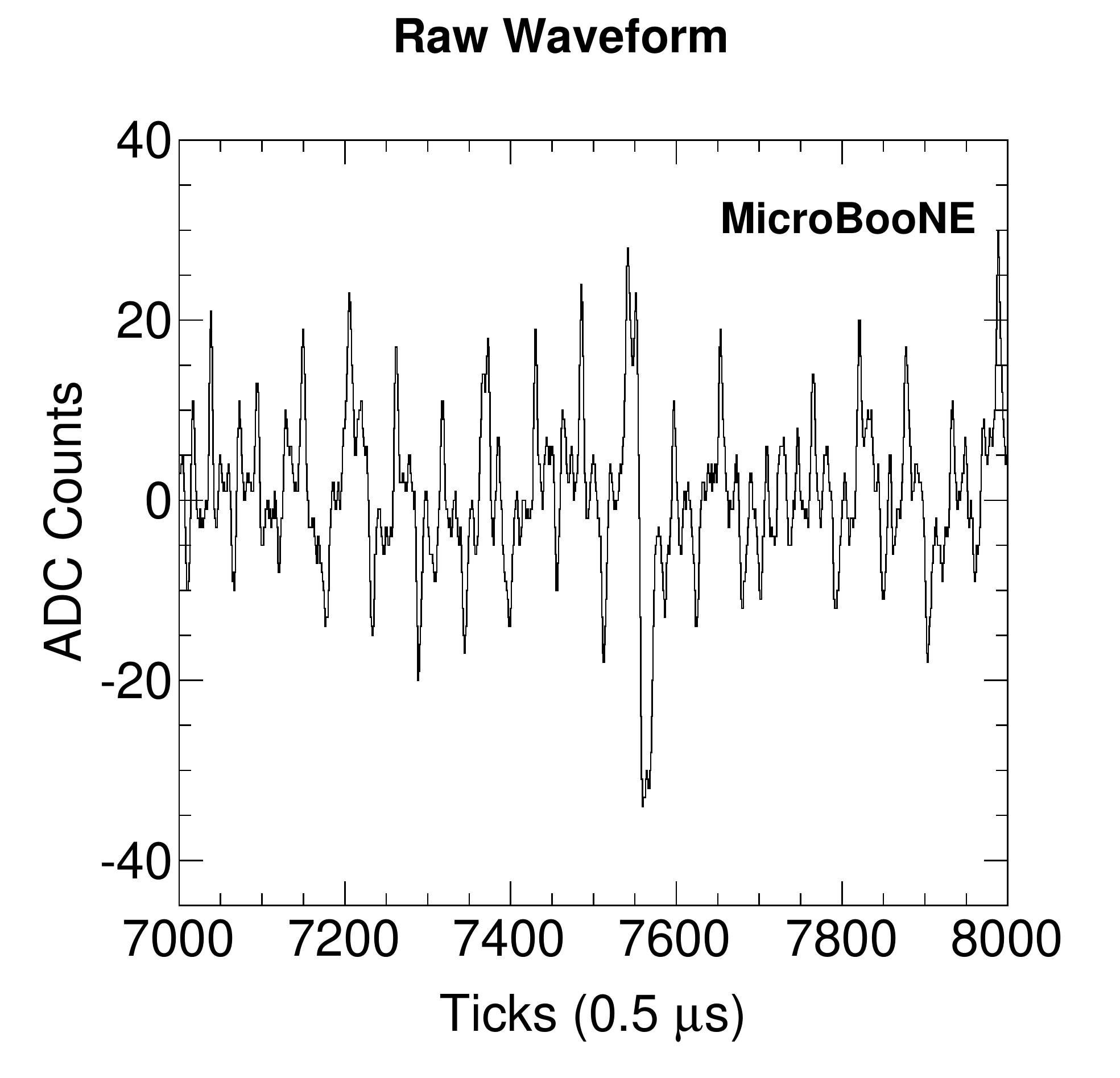} 
\label{freq_1}
}
\subfigure[]{
\includegraphics[width=\fighalfwidth]{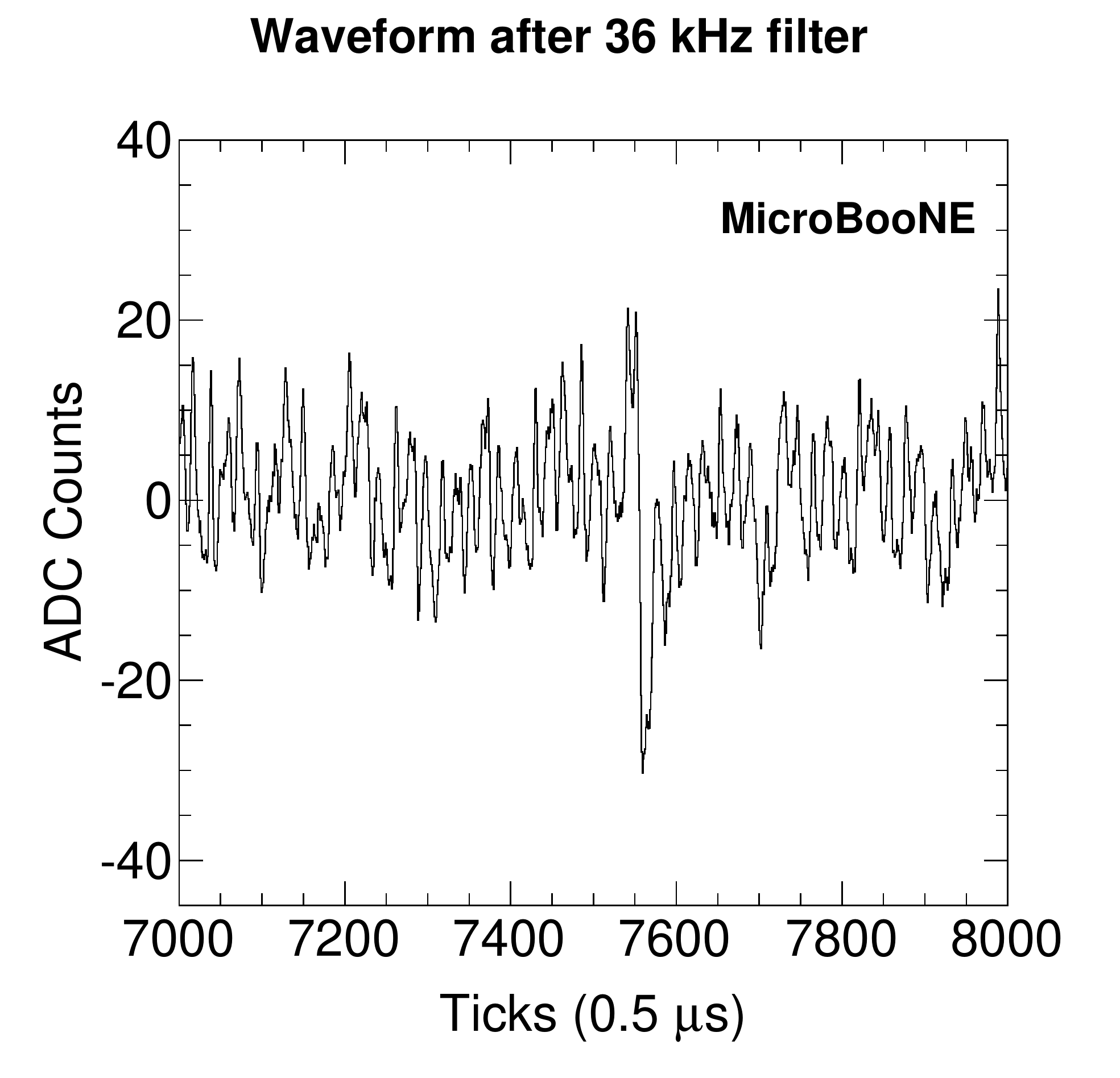} 
\label{freq_2}
}
\subfigure[]{
\includegraphics[width=\fighalfwidth]{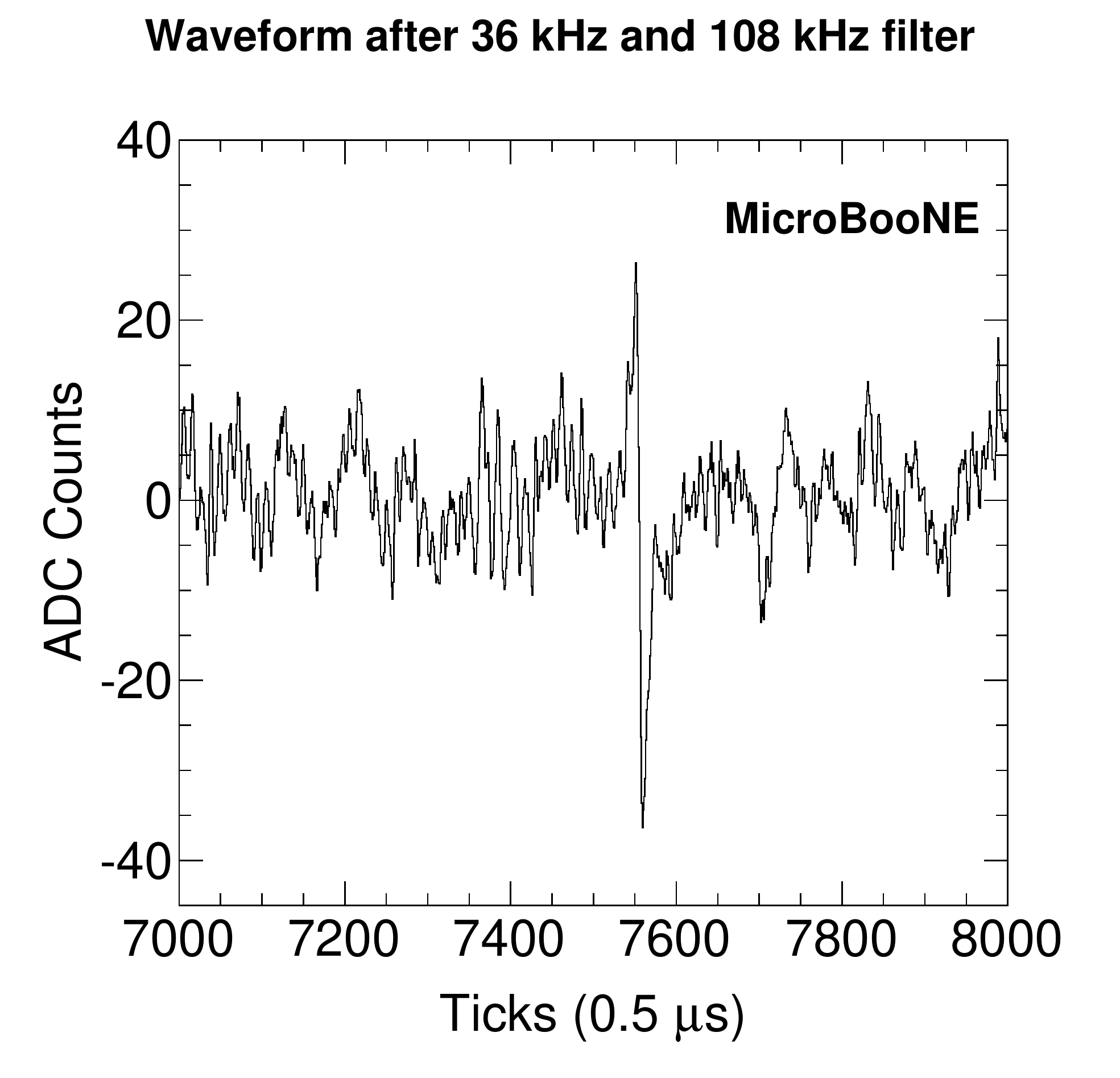} 
\label{freq_3}
}
\subfigure[]{
\includegraphics[width=\fighalfwidth]{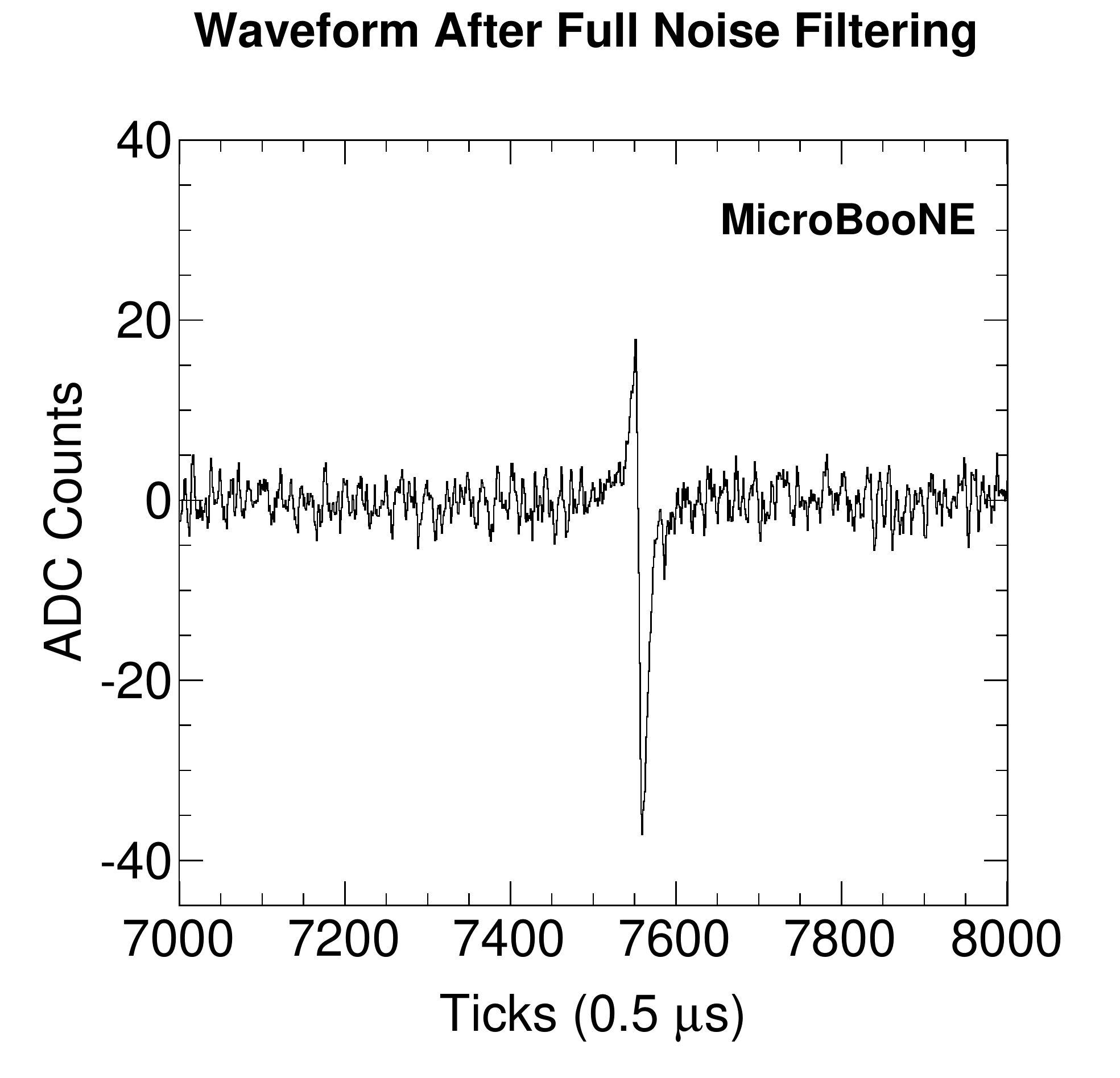}
\label{freq_4}
}
\caption[freq_filt]{(a) A U plane raw waveform from data, (b) the same
  waveform after filtering out \SI{36}{\kHz} noise, (c)
  the same waveform after filtering out both \SI{36}{\kHz} and
  \SI{108}{\kHz} noise components, and (d) the
  waveform after the full noise filtering chain.}
\end{figure}

Figure~\ref{freq_4} shows an example
waveform after this coherent noise removal by offline noise filtering.
Section~\ref{sec:hardware}
describes how this coherent noise from the voltage regulators has been
largely suppressed with a subsequent hardware upgrade of the service boards.

\subsection{The HV power supply noise}
\label{sec:harmonic}

As shown in figure~\ref{excess_1}, a series of narrow
frequency lines appear at odd harmonics of \SI{36}{\kHz} which 
correspond to the ripple frequency of the HV power supply.  The two
highest peaks are near \SI{36}{\kHz} and \SI{108}{\kHz}, and other smaller
harmonics are evident.  The peak magnitude of this \textit{harmonic
  noise} collected by channels of the first induction U plane is 
  $\approx$15 ADC counts. The noise is attenuated by about a factor of three for the
V plane channels due to shielding by the U plane wires and becomes negligible for the Y plane channels due to
shielding by the other two wire planes.

A simple estimation for the dependence of the charge at the anode plane
to tiny variations in the cathode potential, \SI{2.5}{\meter} away, was
made. For a \SI{2.5}{\meter} wire length with a capacitance to the cathode of
\SI{\approx20}{\fF/\meter},\footnote{\SI{\approx20}{\fF/\meter} is the estimated value of the single wire capacitance to the cathode and is based on a planar capacitance calculation between the cathode and anode. This calculation does not include the shielding effect by the field cage, which makes the actual value smaller.}
the induced charge would be \SI{\approx0.05}{\fC/\mV}, resulting in
\num{\approx300} $e^-$\si{/\mV}. The observed peak amplitude of the waveform is
\num{\approx15} ADC counts which corresponds to a charge of \SI{\approx0.5}{\fC}. 
Therefore, a variation as small as \SI{\approx10}{\mV} at the cathode can produce the
observed noise. This corresponds to about five parts in $10^8$ of the drift voltage.

An offline filter directly removes this harmonic noise by masking
discrete bins in the frequency domain. Figure~\ref{freq_1} has no filtering applied and the bipolar
signal is almost indistinguishable from noise. Figure~\ref{freq_2} and~\ref{freq_3} 
show the same waveform
after filtering out the \SI{36}{\kHz} and \SI{108}{\kHz} noise,
after which the characteristic bipolar induction signal can be more
clearly identified. While some substantial noise remains, this is
mitigated by an offline noise filter as described in section~\ref{sec:zig-zag}.

\subsection{\SI{900}{\kHz} burst noise}
\label{sec:zig-zag}

\begin{figure}[!h!tbp]
\center
\subfigure[]{
\includegraphics[width=\figwidth]{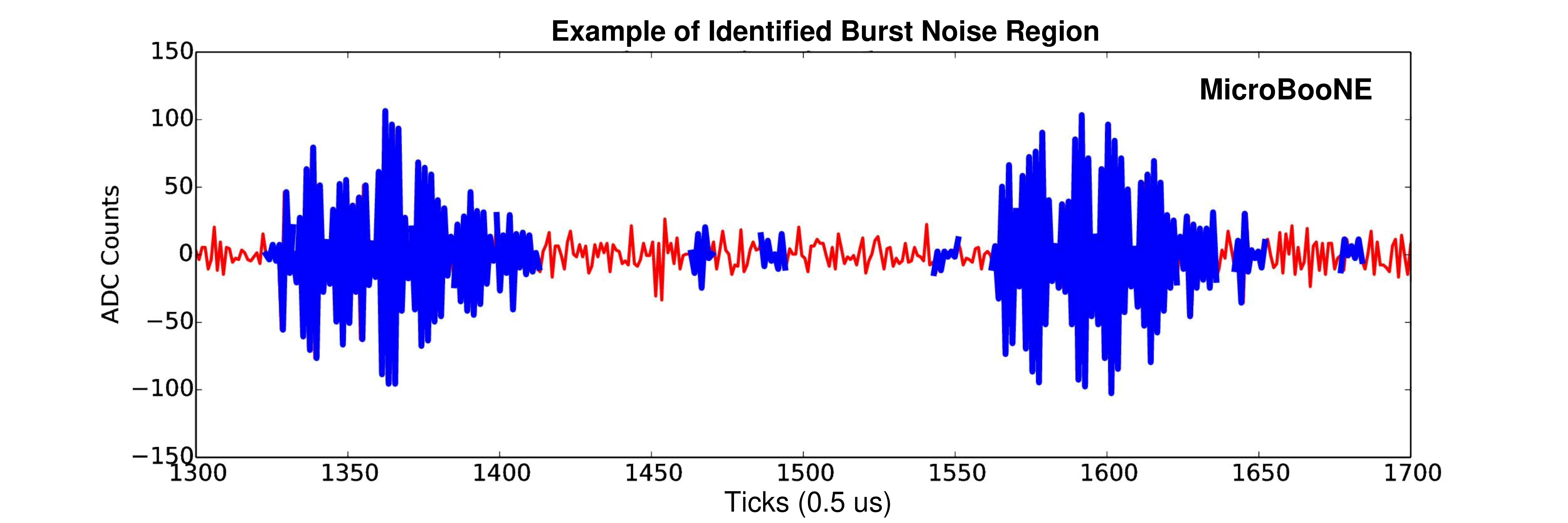}
\label{zigzag_1}
}
\subfigure[]{
\includegraphics[width=\figwidth]{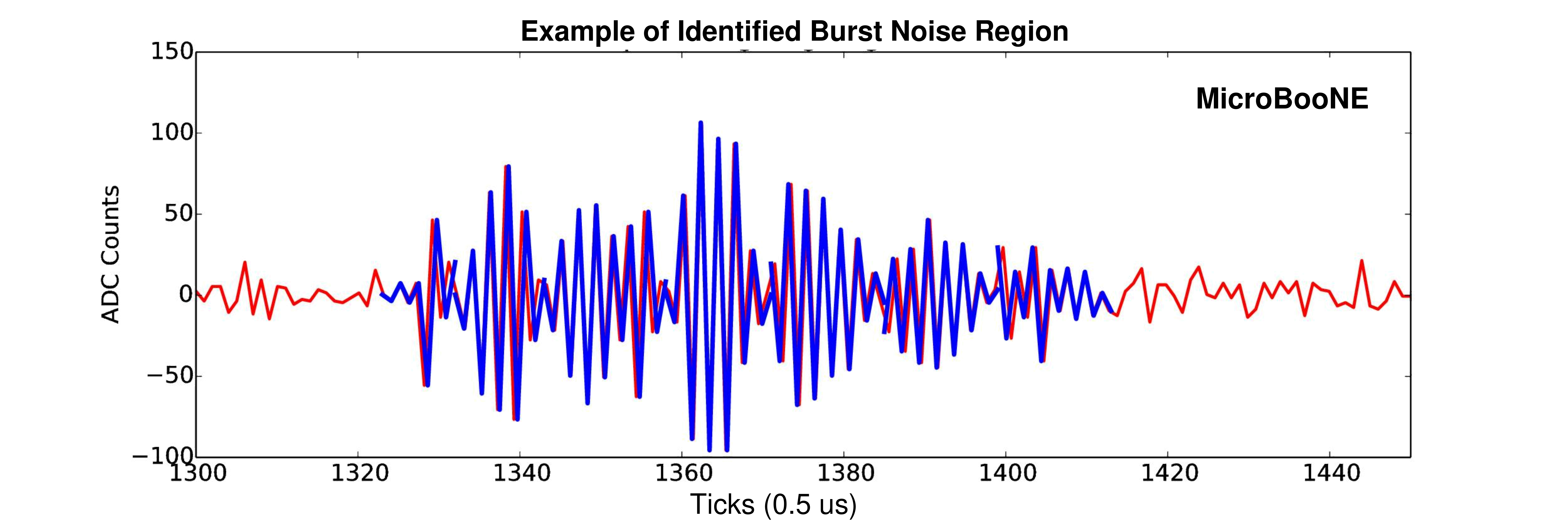}
\label{zigzag_2}
}
\caption[zigzag_time_domain]{(a) A V plane waveform showing an example occurrence of the ``\SI{900}{\kHz} burst noise''.  The regions identified as exhibiting a burst of noise are highlighted in blue, and (b) zooms into the first of the two large bursts regions shown in the top panel.}
\end{figure}

The least problematic excess noise is the so called ``\SI{900}{\kHz} burst noise''. 
A waveform showing an example of this noise
is shown in figure~\ref{zigzag_1} and~\ref{zigzag_2}.  

The measured period of the oscillation (or ringing) seen in these
bursts is \SI{1.1}{\us}, as anticipated given the approximately
\SI{900}{\kHz} peak seen in the frequency spectrum in
figure~\ref{excess_1}.  The bursts are intermittent and have
amplitudes that vary over time. This noise has been observed in data
from wires that are located at one corner of the detector.

No mitigation of this noise is performed as it is naturally
attenuated to acceptable levels by the anti-alias filter with a
\SI{2}{\us} peaking time.  This can be seen in a relative manner by
comparing the spectra shown in figure~\ref{excess_1} and~\ref{excess_2}. If
this noise increases in the future, it can be
mitigated with a simple low pass offline filter.
However, in the case of the few misconfigured channels described in
section~\ref{sec:misconfig}, the data is taken with a \SI{1}{\us} peaking
time and figure~\ref{excess_1} shows a prominent peak at
\SI{900}{\kHz}. In the offline filter, the \SI{1}{\us} shaping introduced by
the amplifier is replaced by a \SI{2}{\us} shaping filter. This
results in the data from the misconfigured channels having effectively
the same high frequency attenuation as the data from the nominally
configured channels.

The exact origin of this burst noise has not been conclusively identified.
However, the power switching circuits related to either the PMT high
voltage power supply or the interlock system power supply are
currently suspected as a possible source.

\subsection{Impact of noise filtering on TPC signal}

In order to examine the performance of the coherent noise filter on the data, a simulated data study was conducted that 
overlays pulse shapes on top of recorded TPC data. The pulse shapes correspond to the average field 
response convolved with the readout electronics response for a given amount of ionization charge on a wire. The
impact of the noise filtering process on these simulated signals is evaluated for signals produced by isolated point-like charges and 
for signals produced by minimum ionization tracks traveling parallel to the wire plane and perpendicular to a given wire. Signals produced by parallel tracks are expected to 
be most affected by the noise filter, as adjacent wires will have common signals that can be misidentified as noise by the 
correlated low frequency noise subtraction algorithm.

\begin{figure}[!h!tbp]
\center
\subfigure[]{
\includegraphics[width=\fighalfwidth]{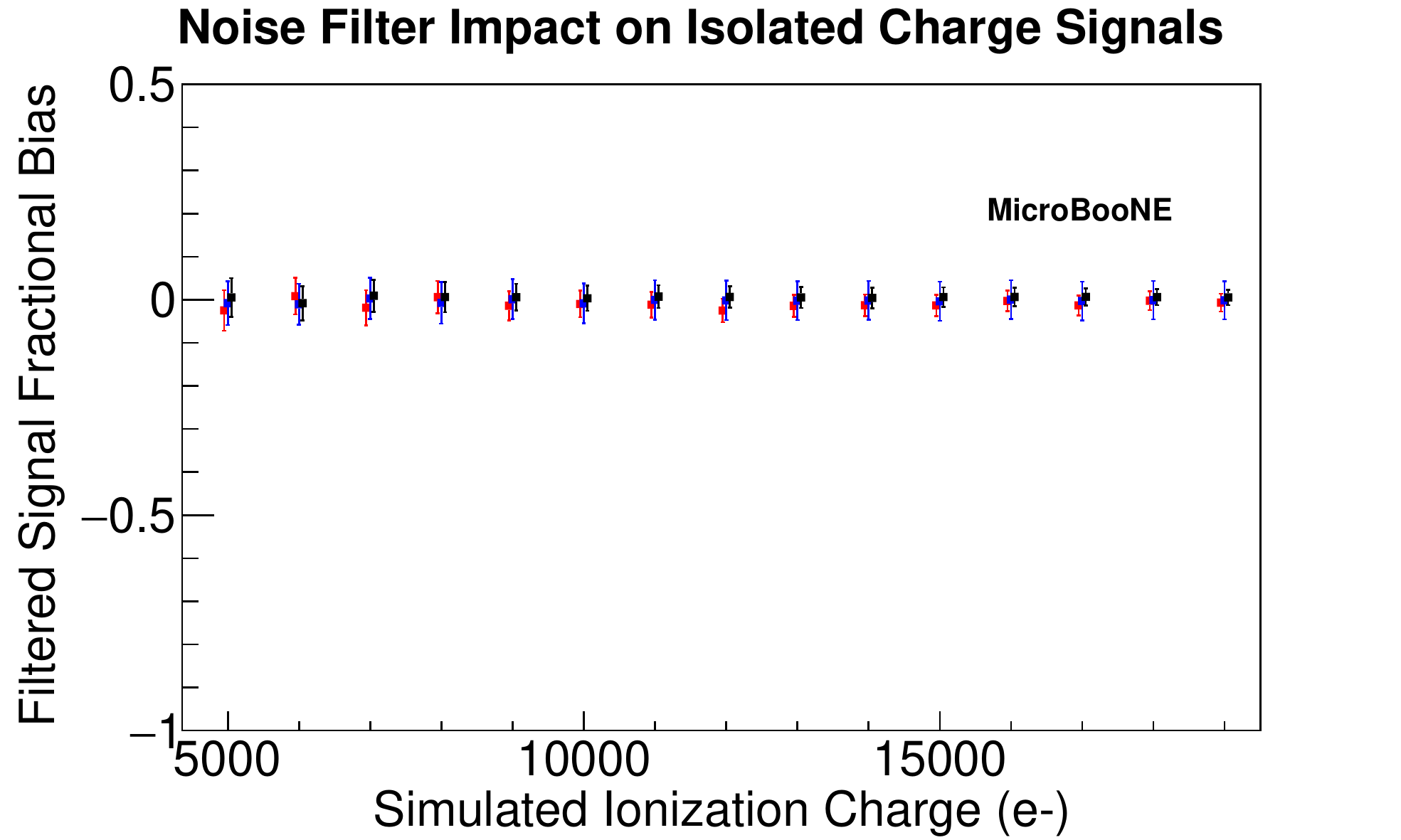}
\label{filter_impact_singleplots}
}
\subfigure[]{
\includegraphics[width=\fighalfwidth]{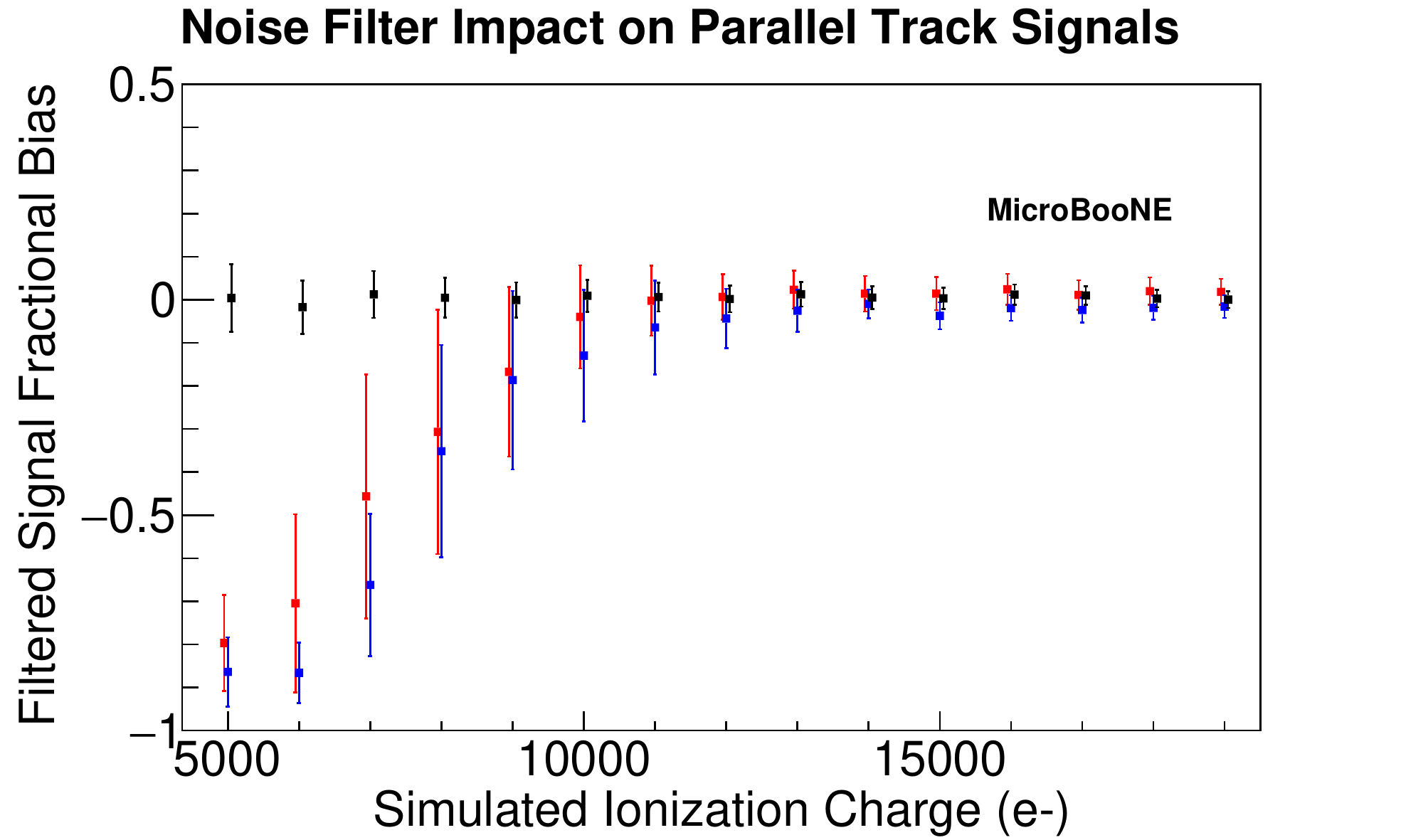}
\label{filter_impact_angularplots_v}
}
\caption[filter_impact_singleplots]{Fractional bias of filtered signals for different amounts of simulated ionization 
charge in the case of (a) isolated charge; (b) charge signals produced by a track traveling parallel to the wire plane for U (red), 
V (blue), and Y (black) plane wires. In the case of (a) isolated 
ionization charge the measured bias introduced by the noise filter is consistent with zero even for very small amounts of charge. 
For charge signals produced by a track traveling parallel to the wire plane and perpendicular to a given wire (b) there is significant bias for small amounts of charge, 
but this bias becomes very small for signal sizes expected for minimum ionizing particles ($\approx$16000 electrons).}
\end{figure}

The impact of the filtering process on ionization charge signals is evaluated by measuring the filtered signal sizes
and comparing them to the expected values. The fractional bias $F_{Bias}$ in filtered signal waveforms is defined as:
\begin{equation}
  F_{Bias} =  \frac{ S_{Filter} - S_{True} }{ S_{True} } ,
  \label{eqn:fracbias}
\end{equation}
where $S_{Filter}$ is the measured filtered signal size and $S_{True}$ is the true value used in the simulation.

The fractional bias in measured signal size as a function of simulated signal size for isolated charge signals is 
summarized in figure~\ref{filter_impact_singleplots}, and shows that the noise filter introduces relatively little 
distortion, even for very small isolated charge signals. Figure~\ref{filter_impact_angularplots_v} show the measured fractional 
bias of filtered signals for tracks running parallel to the anode plane and perpendicular to a given wire. Small signals are distorted by the correlated noise subtraction 
resulting in a large fractional bias, however this effect is very small for signal sizes expected from minimum ionizing particles ($\approx$16000 electrons). 
Even for large signals, U-plane signals have a $\approx$2.5 $\%$ negative fractional bias due to attenuation of the long rising edge by the noise subtraction algorithm. 
Following the service board upgrade, the fractional bias for small filtered signals was significantly reduced.

\section{Residual noise levels}
\label{sec:RMS}

\begin{figure}[!h!tbp]
\center
\includegraphics[width=\figwidth]{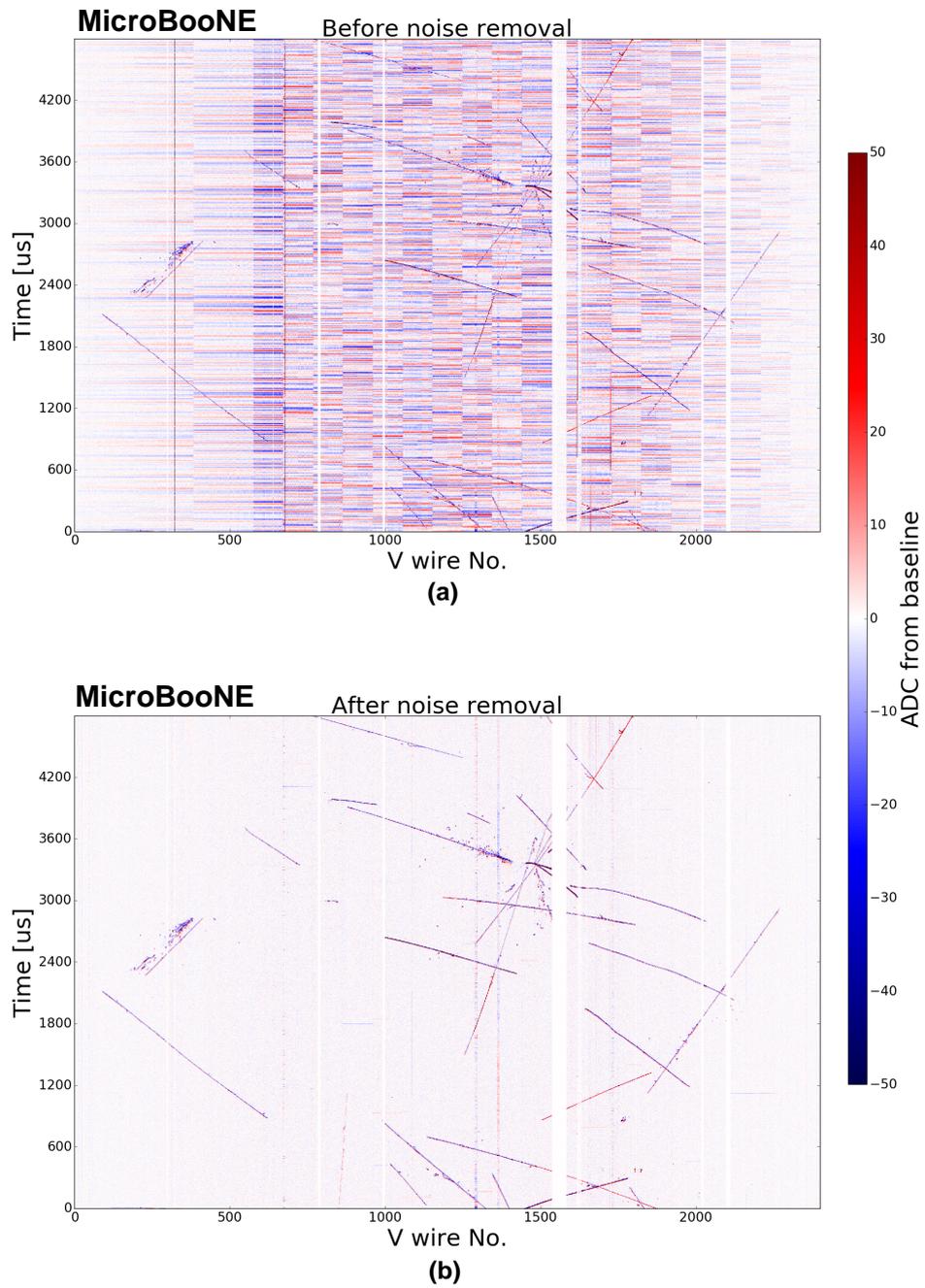}
\caption[Event display]{2-D event display of the V plane from run 3493 event 41075 showing the raw signal (a) before and (b) after offline noise filtering. A clean event signature is recovered once all the identified noise sources are subtracted.}
\label{evt_dis2}
\end{figure}

An example waveform from before and after the full noise filtering
procedure is shown in figures~\ref{freq_1} and~\ref{freq_4}. 
The noise level for this example waveform is reduced by a
factor of \numrange{3}{5} for the induction planes and a factor of 
\numrange{2}{3} for the
collection plane. To give a broader picture of the effect of the noise
filtering across an entire event, an example two-dimensional (2D) event display for the V plane is shown in
figure~\ref{evt_dis2}. The noise filtering removes most
of the excess noise while preserving the signal.

\begin{figure}[!h!tbp]
\center
\subfigure[]{
\includegraphics[width=0.50\figwidth]{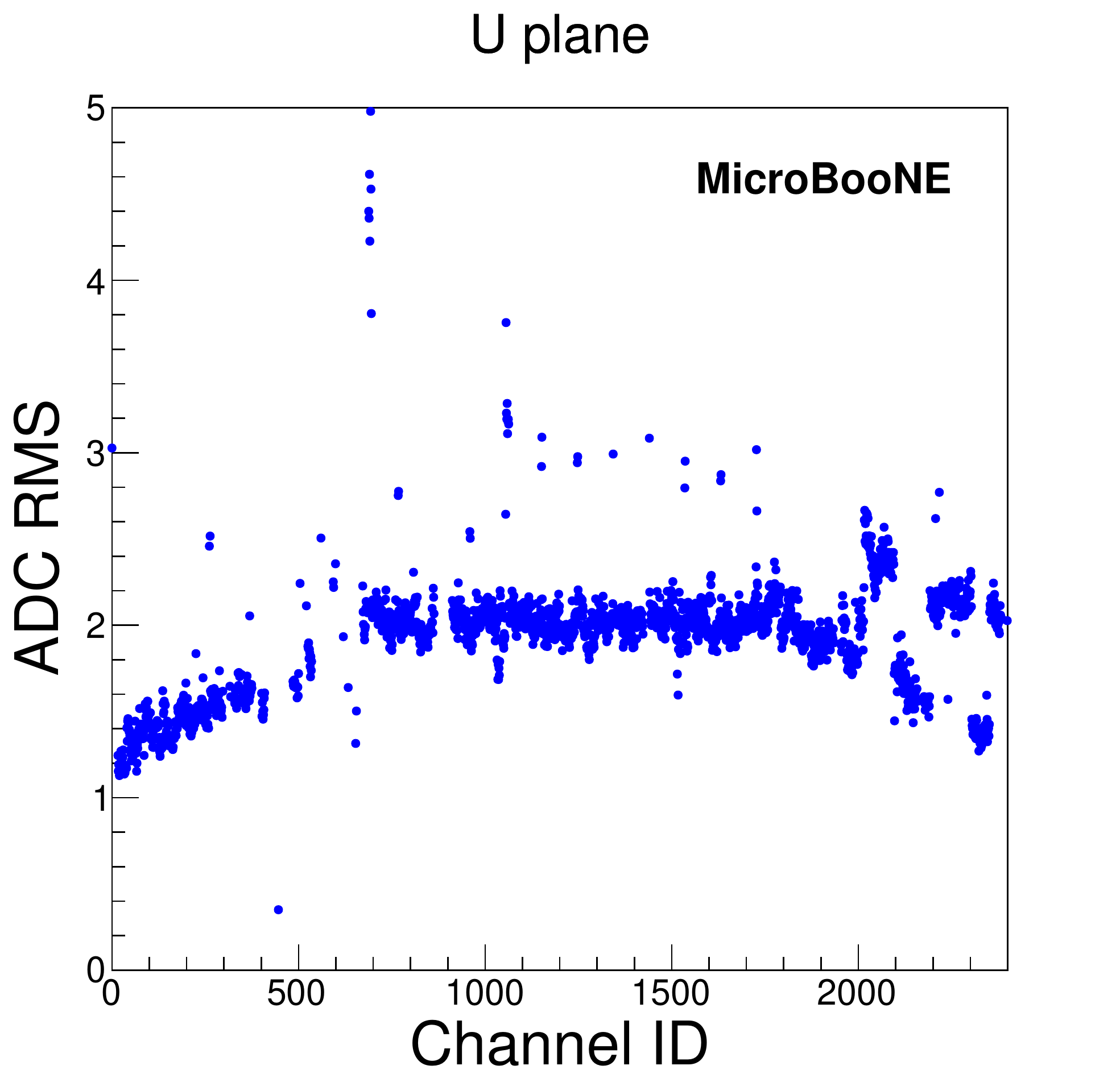}
}
\subfigure[]{
\includegraphics[width=0.50\figwidth]{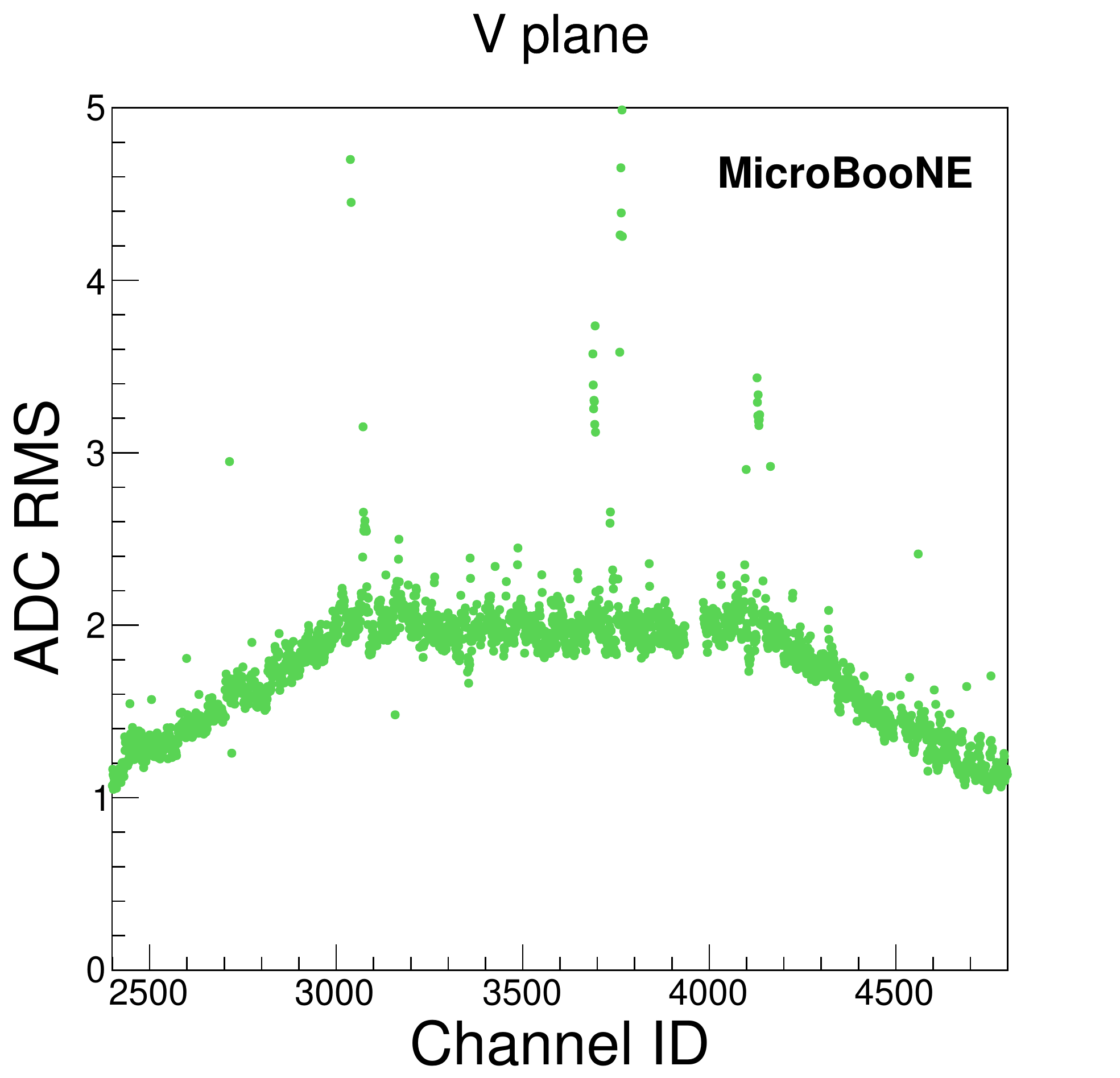}
}\\
\subfigure[]{
\includegraphics[width=0.50\figwidth]{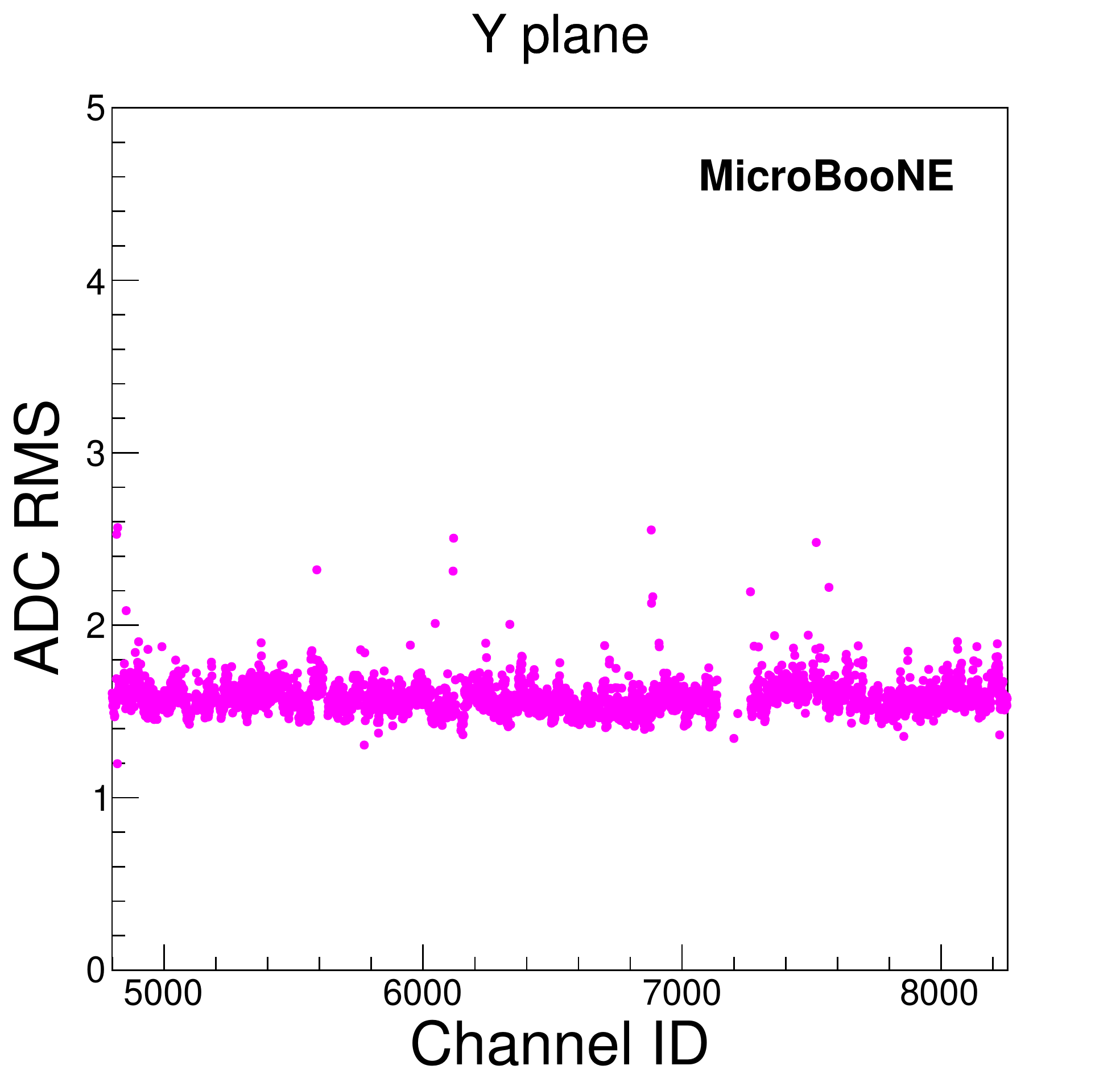}
}
\caption[ADC RMS all channels]{The residual noise level after
  filtering in terms of the ADC RMS value for all channels (excluding
  $\approx$860 non-functioning channels) from run 3455 event 6 is shown for (a) U,
  (b) V, and (c) Y planes. High RMS bands in the U plane (a) correspond to
  misconfigured channels as described in
  section~\ref{sec:misconfig}. These channels are associated with
  larger noise as the \SI{1}{\us} peaking time setting leads to $\approx$10\%
  higher inherent noise than the \SI{2}{\us} peaking
  time setting and the lowest gain setting leads to a much larger
  contribution of the electronics noise beyond the preamplifier.}
\label{rms_before}
\end{figure}

Figure~\ref{rms_before} shows the residual noise in terms of the ADC RMS
value after the full noise filtering is performed. 
%
%
%
There are
\num{\approx860} channels not shown here that were identified
as non-functioning as described in section~\ref{sec:dead}. The remaining
channels with anomalously high ADC RMS are each examined in both the
time and frequency domain. Although the observed noise frequency
spectra on these channels are different from the majority of ``live''
channels, good induction signals can still be seen. Therefore, we keep
these channels (\num{\approx84} channels in run 3455) in the data analysis.

\begin{figure}[!h!tbp]
\center
\includegraphics[width=\figwidth]{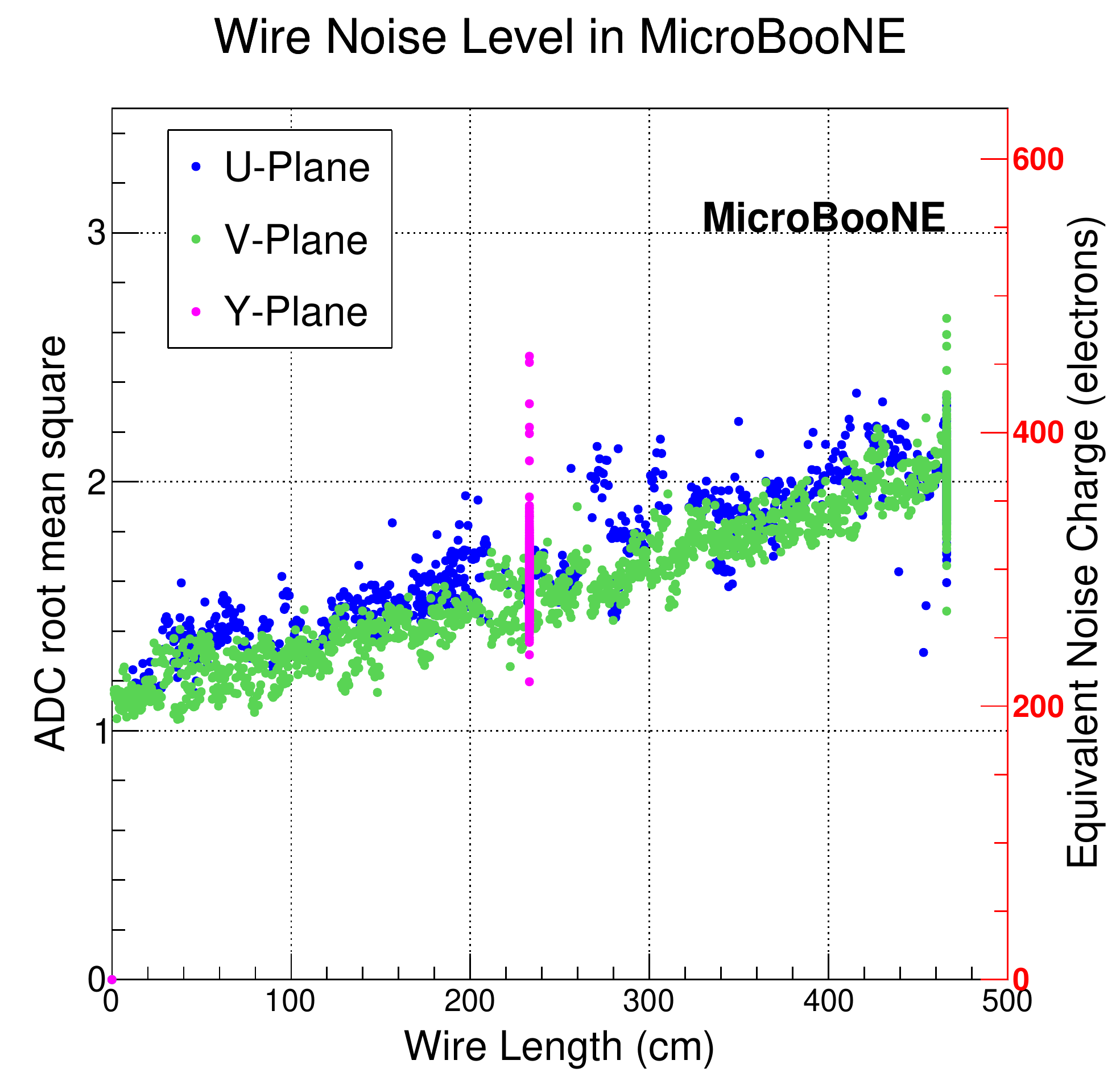}
\caption[Residual noise vs wire length]{Noise level in terms of the ADC RMS value and ENC are plotted with respect to the wire length. The vertical band in magenta is all of the Y-plane wires which have the same length. The vertical green and blue bands at the right most end of the plot contain results from the longest (\SI{4.7}{\meter}) wires in the U and V planes which are of equal length.}
\label{rms3} 
\end{figure}

As explained in section~\ref{sec:RC}, the amount of electronics noise is
expected to have a dependence on the wire length. Figure~\ref{rms3}
shows the magnitude of the electronic noise with respect to wire
length. Fitting the data with equation~\ref{eqn:rms_fit1}, gives best fit
parameters of $x = 0.90$ ADC, $y = 0.79$ ADC and $z = 0.22$
ADC/m. Wire capacitance can be calculated using figure~\ref{fig:enc_cin},
which shows the variation of ENC with input capacitance, and from
these fit parameters. The wire capacitance is found to be \SI[separate-uncertainty = true]{23 \pm 1}{\pF/\meter}, 
with the two assumptions that the ENC contribution 
from ADC digitization is $\approx$150 electrons and the simulated ENC vs. input capacitance 
curve is solely due to the intrinsic noise. This value is within 10\% of the estimations
given in section~\ref{sec:cold_noise}.

The noise level in terms of ADC is converted to the equivalent noise
charge (ENC) as follows:
\begin{equation}
  \frac {\rm{ENC}} {\rm{ADC}} =  \frac{\SI{2000}{\mV}}{\SI{4096}{ADC}} \times \frac{\SI{1}{\fC}}{\SI{14}{\mV}} \times \frac {1}{1.2}\times \frac{{6241~e^-}}{\si{\fC}} = 182~\frac{e^-}{\rm{ADC}}.
  \label{eqn:adctoenc}
\end{equation}

The gain conversion factor given in equation~\ref{eqn:adctoenc} is consistent 
with test stand and in-situ pulser measurements. The factor of \SI{14}{\mV/\fC} is the gain of the cold ASIC, and
1.2 is the gain of the intermediate amplifier and receiver/ADC board. 
As discussed in section~\ref{sec:impulse}, induction plane channels with 
baseline of \SI{900}{\mV} have $\approx$3\% lower gain, i.e. $\approx$3\% higher 
gain conversion factor than the collection plane channels with baseline of \SI{200}{\mV}.
Therefore, the nominal ENC
for the longest U plane wires is \num{\approx400} electrons, longest V plane wires 
is \num{\approx380} electrons and Y plane wires is \num{\approx300} electrons.

\begin{figure}[!h!tbp]
\center
\includegraphics[width=\figwidth]{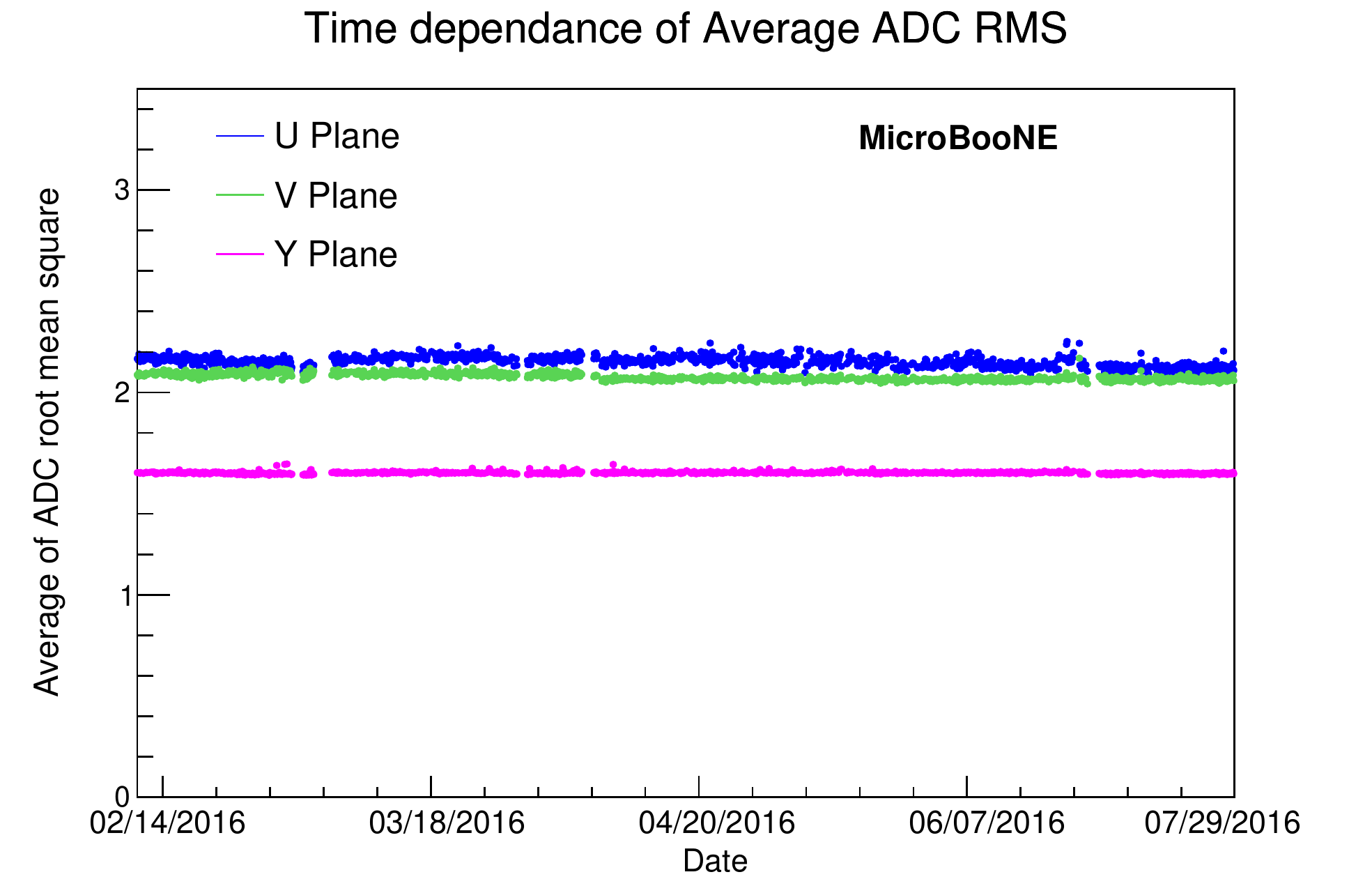}
\caption[RMS_Run]{Average of ADC RMS for constant length U, V and Y wires as a function of time. The average is taken over ten events per run. From event to event, there could be fluctuation as large as 
$\approx$0.1 ADC or 14 electrons. This plot covers the data taking period from mid February to the end of July 2016, before the summer accelerator shutdown.}
\label{rms_run} 
\end{figure}

Figure~\ref{rms_run} shows the stability of the ADC RMS for
same-length wires (longest wires for induction planes and all
collection plane wires) over the run period from February through
July, 2016.  Each point in the plot is the average over ten events from a 
given run. From event to event, there could be fluctuation as large as 
$\approx$0.1 ADC, equivalent to 14 electrons. 

\section{Peak-signal to noise ratio}
\label{sec:PSNR}

\fixme{It would be useful to have an example induction waveform that describes the PSNR definition.}

As charge drifts near the sense wires, the currents it induces provide
a measure of its distribution in space and time. This measure is
distorted by the nearby electrostatic field of the detector. 
In particular, the bipolar waveforms recorded on the
induction plane channels are not a direct measure 
of the original charge distribution.

Independent of this, the amount of signal measured for a given charge
distribution depends on the location and orientation of its
distribution in the detector volume.  Effects of diffusion and
electron absorption in the LAr will reduce the measured signal as the
drift distance increases.  In addition, signals, particularly on the induction
planes, vary substantially depending on the relative angle between 
a given ionization track and a sense wire.

On the other hand, the excess and inherent noise are largely
independent of the electrostatic field distortion and the ionization
topology.  As a consequence, any relative measure of signal
and noise is non-trivial and likely must be constructed under various
simplifying assumptions. ENC relates noise to the 
charge collected in a very short time (delta function).
As such it can not be used to express noise
relative to signals from induction wires. The requirement of a point
charge of known magnitude for the signal is not possible to produce
with events available in MicroBooNE data. An alternative metric that trades these limitations for others is
presented here.  The ``peak-signal to noise ratio'' (PSNR) takes
for its charge distribution an ideal MIP track (``standard candle'')
running near and parallel to the wire planes and perpendicular to a
given wire.  This removes the topological variance described above.
The ``peak-signal'' numerator is the number of electrons taken 
as the peak ADC sample.  In the case of
collection signals, this peak is a good measure of the number of
nearby electrons.  For induction plane signals it is only a partial,
truncated measure due to the bipolar detector field response.  The
``noise'' denominator of the ratio is simply the ADC RMS sampled
outside the signal region of a given waveform.

Although the ideal standard candle is closer to the real charge
distributions seen in MicroBooNE than is a point charge, statistically
there are very few tracks that exactly meet this condition. To increase the sample
size, tracks which otherwise come close to meeting the standard candle
criteria, are allowed to pass by wires at some angle $\beta$ which is
not perfectly perpendicular. The resulting signal is then corrected
for the geometrical factor given their reconstructed track direction.

\begin{figure}[!tbp]
\centerline{
  \includegraphics[width=\figwidth]{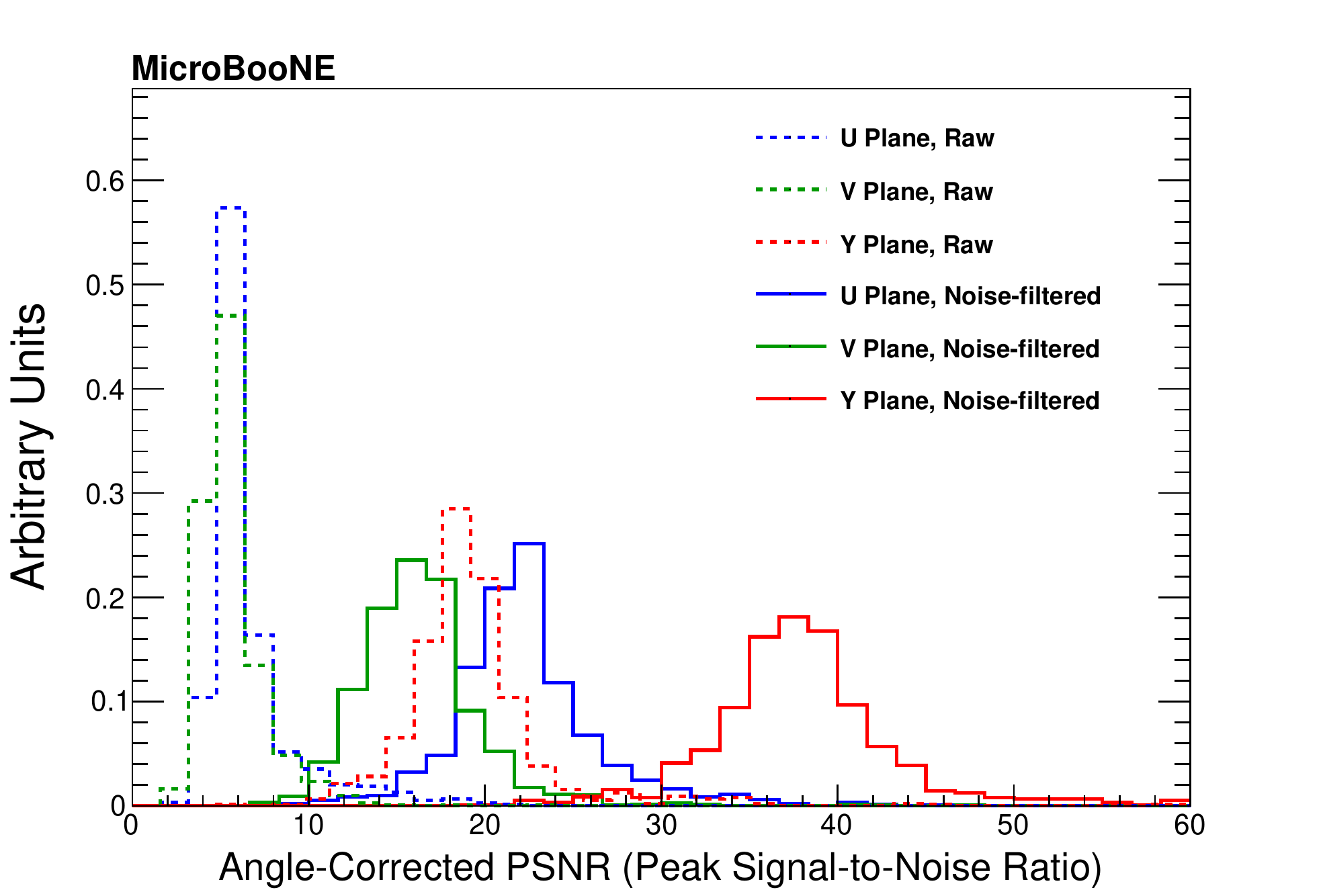}
}
\caption[SNplots]{Angle-corrected peak-signal to noise ratio (PSNR)
  from a ``standard candle'' signal as described in the text before (dashed) and after (solid) offline
  noise filtering for channels in the U (blue), V (green) and Y (red)
  planes.  Significant improvement is seen in the PSNR after the
  application of noise filtering due to the reduction of
  noise levels.  }
\label{SNplots}
\end{figure}

With this definition and correction in mind, the number of
ionization electrons arriving at a single wire for the selected events
can be estimated as

\begin{equation}
  N_{e^-} = \frac{\frac{dE}{dx} \cdot \Delta L}{W_{\mathrm{ionization}}} \cdot \frac{\epsilon_{\mathrm{quenching}}
  \cdot \epsilon_{\mathrm{lifetime}}}{\sin(\beta)},
\end{equation}

where the most probable energy loss $\frac{dE}{dx}$ is 
\SI{1.8}{\MeV/\cm} with $\approx$8\% fluctuation near the peak. 
In this case, the travel distance $\Delta L$ is the wire pitch
(\SI{3}{\mm}). $W_{\mathrm{ionization}}$ is the average energy to
generate an ionization electron in LAr and is assumed to be
\SI{23.6}{\eV}~\cite{lar_property}.  The charge recombination
factor $\epsilon_{\mathrm{quenching}}$~\cite{Acciarri:2013met} and
the attenuation due to finite electron lifetime
$\epsilon_{\mathrm{lifetime}}$ are assumed to be 0.7 and 1.0,
respectively.
For a unit geometrical correction ($\sin(\beta) = 1$)
the final number of ionization electrons
arriving at a wire is $N_{e^-} \approx \num{16000}$.
Note, that the intrinsic fluctuation of energy loss per unit track
length for a MIP particle inside LAr is $\approx$8\%.
\fixme{This equation is not used for anything, is it?}

Figure~\ref{SNplots} shows the PSNR for all three planes before and
after the noise filtering. The calculation of PSNR uses a
sample of $t_0$-tagged cosmic muon tracks,\footnote{The time ($t_0$) with respect to the zero time of the data acquisition system at which anode-piercing through-going cosmic ray muons enter the MicroBooNE TPC.} near the anode plane in events collected using external triggers. 
Tracks traveling in a direction within \SI{20}{\degree} of
parallel to the anode plane are selected and the $\beta$ angular correction to
their signal is made.  Each track then contributes one
entry per plane to the plot using the median PSNR value associated
with the track in the region of interest.

\begin{table}[tb]
\center
 \caption[PSNR Comparison]{Summary of the mean angle-corrected
    peak-signal to noise ratio (PSNR) for each plane, both before and
    after noise filtering using a ``standard candle'' signal
    as described in the text. }
     \vspace{2mm}
  \label{tab:PSNRcomp}
\begin{tabular}{|l|c|c|c|}
\hline
\bf{Waveform Type} & \bf{U Plane PSNR} & \bf{V Plane PSNR} & \bf{Y Plane PSNR} \\ \hline
Raw Data  & 6.6 & 5.7 & 19.5 \\
After Noise Filtering & 22.3 & 16.2 & 37.9 \\
\hline
\end{tabular}
\end{table}

Table~\ref{tab:PSNRcomp} summarizes the average PSNR values for each
plane before and after noise filtering.  It is clear that the PSNR
for the collection plane is much higher than those of the induction
plane due, in part, to their bipolar field response, as shown in
figure~\ref{fig:tpc_signal}.  However, it is important to point out that
the standard candle selection criteria tends to maximize the
peak-signal of induction plane waveforms.  Charge distributions from
ionization tracks which point approximately normal to the wire planes
will induce a signal which is prolonged in time. The bipolar response
of a nearby induction plane wire will effectively cancel the
middle part of the waveform and give only appreciable sample values
for the portion of the waveform corresponding to the beginning and end of
the ionization track.  Therefore, having the lowest inherent
electronic noise possible through the use of the cold front-end
electronics is crucial to properly reconstruct the ionization charge
information from the induction wire planes.

It is important to mention that while the PSNR metric is useful to
quantify relative improvement in finding small ionization signals as a
function of excess noise mitigation, it is not the most ideal metric
to use to quantify improvement in the resolution of charge estimation
for signals on the wires.  This latter quantity is intimately tied to
our capability to perform precise calorimetric measurements with
ionization signals in the TPC.  In order to determine how noise
filtering influences charge resolution, further digital signal
processing (e.g. deconvolution of electronics shaping and wire field
response) is needed.  Discussion of the digital signal processing
chain at MicroBooNE is outside of the scope of the present work and
will be presented in a future publication.

\section{Hardware upgrades}
\label{sec:hardware}

As discussed previously, excess noise was identified during initial
operations at MicroBooNE and mitigated via offline noise filtering
methods. Subsequent to the run period covered above, and during the
shutdown of the Fermilab accelerator complex over the summer of 2016,
we performed two hardware upgrades an effort to further suppress
the excess noise. These improvements
reduced the amount of filtering required by the offline noise filter and
allowed for lower zero-suppression thresholds to be used by a 
continuous readout system. At this time, no
upgrade has been performed to mitigate the \SI{900}{\kHz} noise for the 
reasons described in section~\ref{sec:zig-zag}.

\begin{figure}[tbp]
\centerline{
\subfigure[]{
  \includegraphics[width=\fighalfwidth]{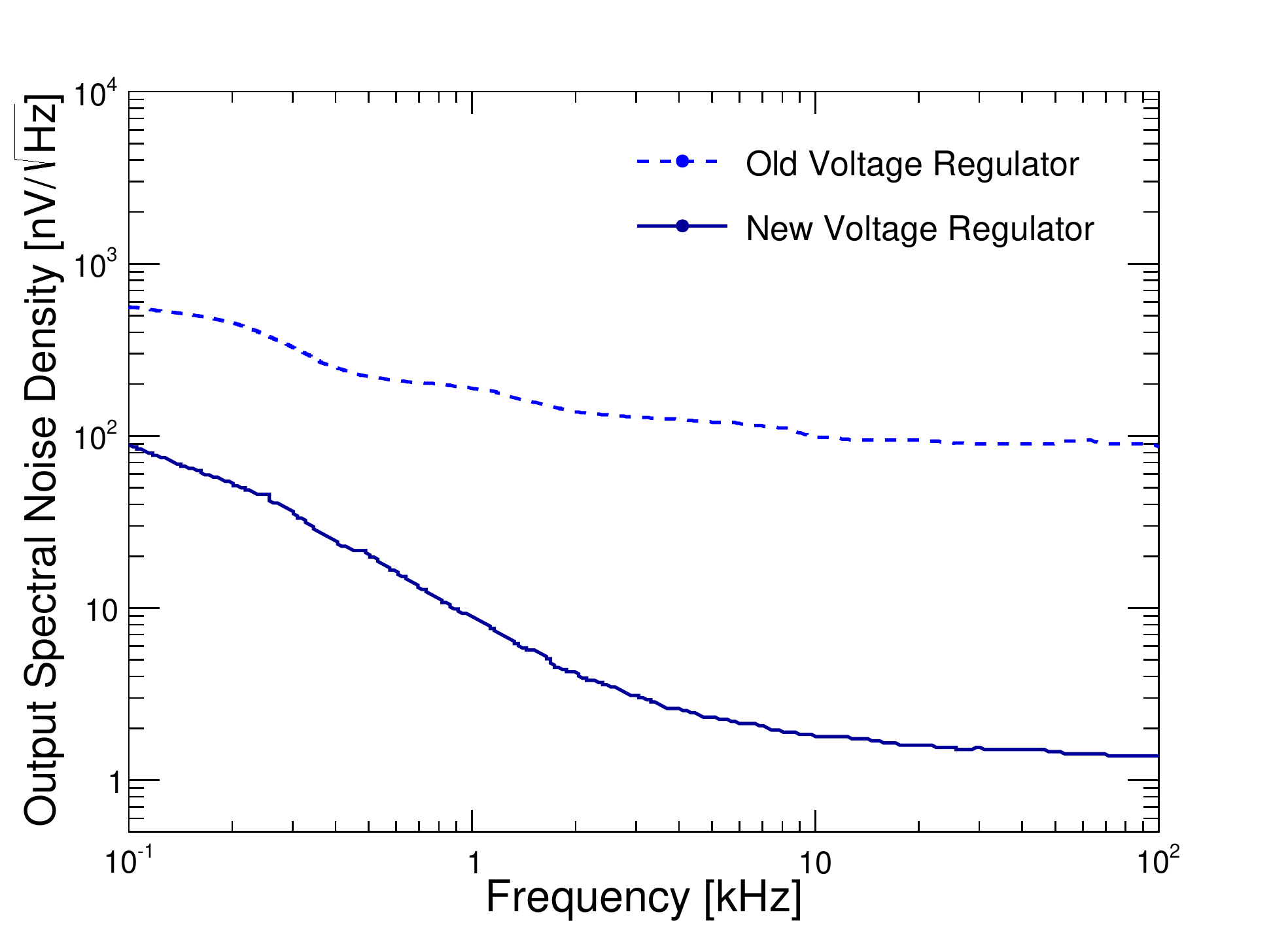}
  \label{HardwareFreqResp-1}
  }
  \subfigure[]{
  \includegraphics[width=\fighalfwidth]{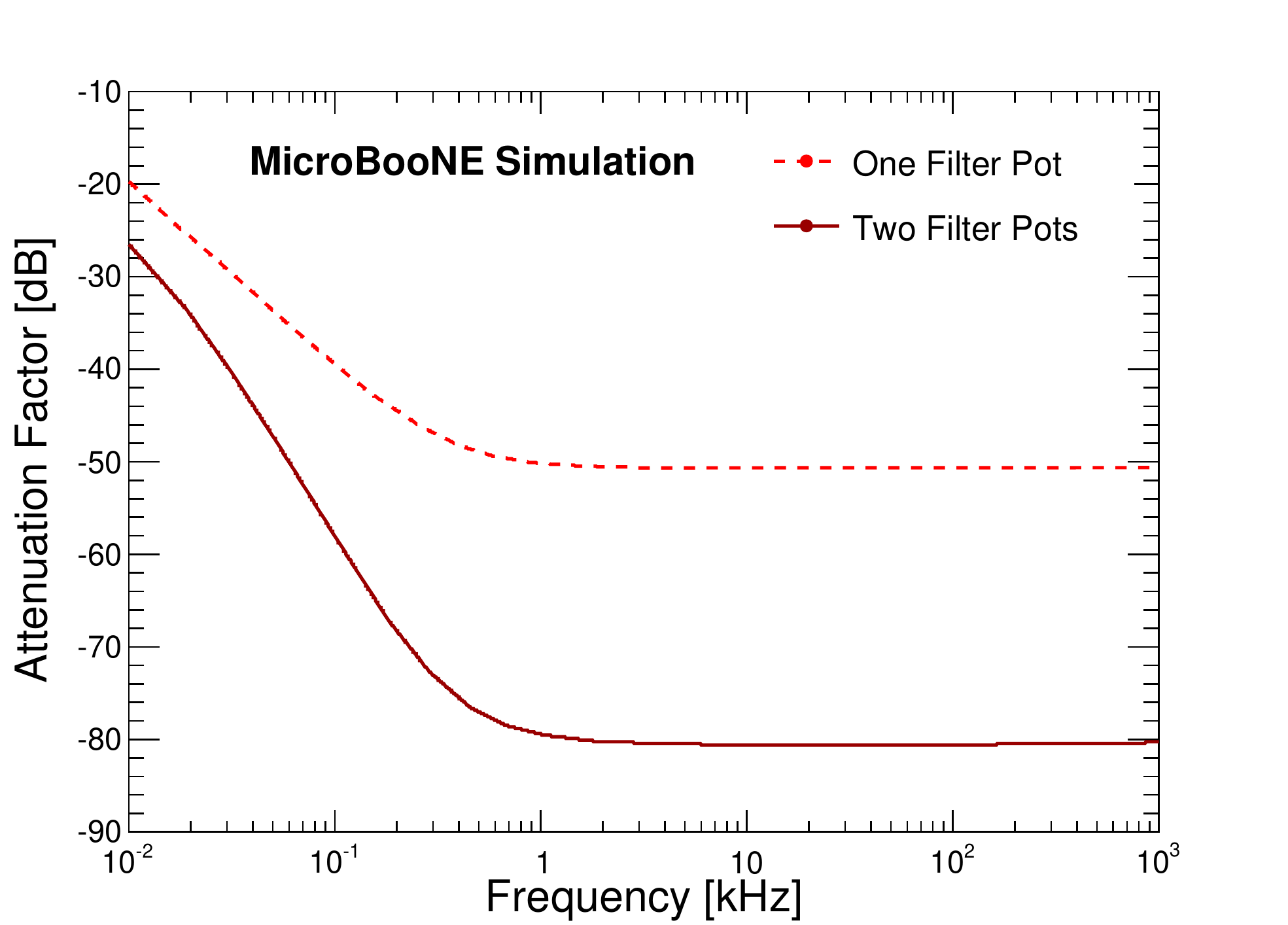}
  \label{HardwareFreqResp-2}
  }
}
\caption[SBFreqResp]{(a) Comparison of spectral noise density for a service 
board with the old voltage regulator, TPS78618, and with a new 
low-noise voltage regulator, ADP7159. (b) Comparison of the
drift HV system frequency response as predicted by a SPICE model for a single 
filter pot versus two identical filter pots.} 
\end{figure}

\begin{figure}[tbp]
\centerline{
  \includegraphics[width=\figwidth]{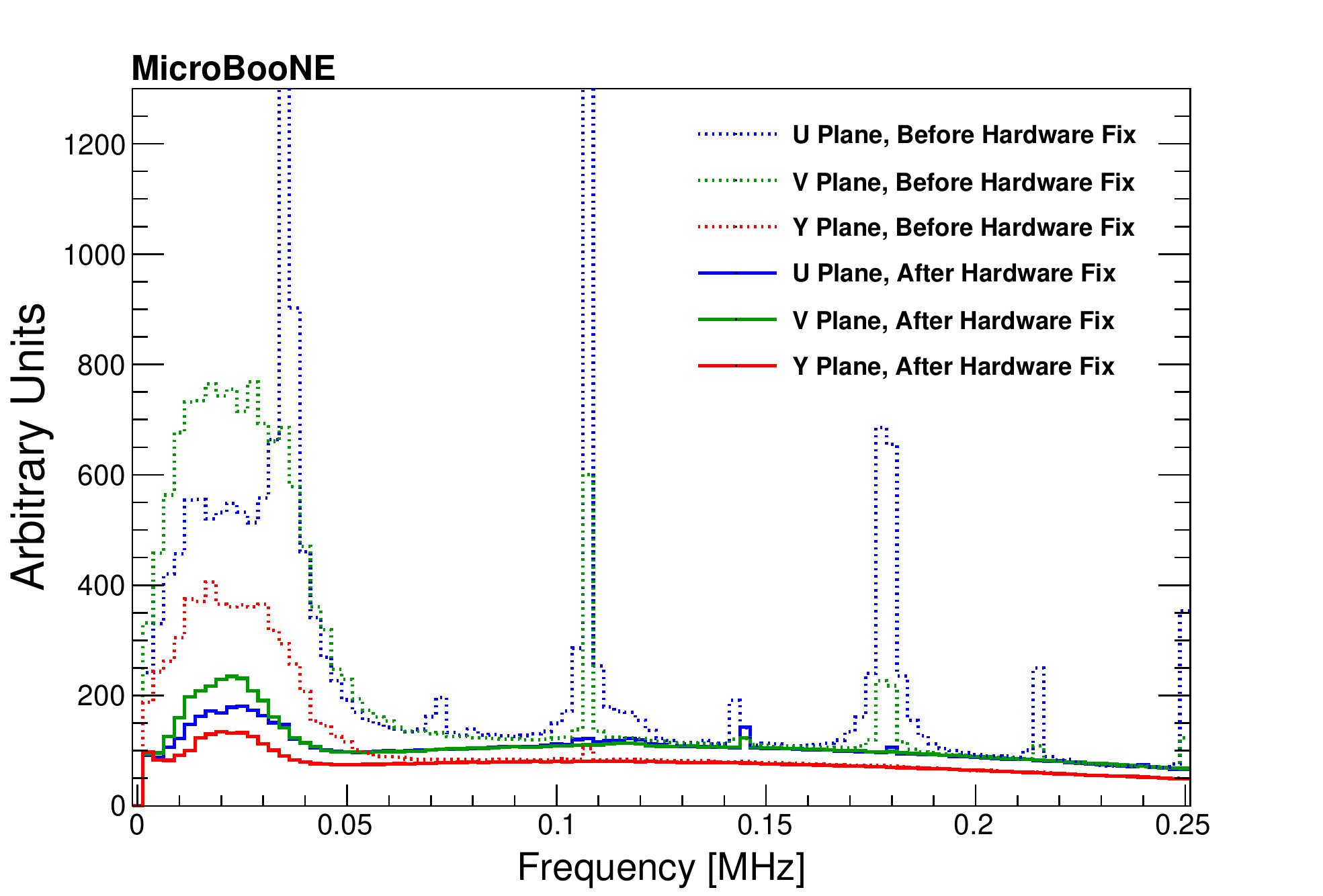}
}
\caption[FFTcomp]{Comparison of the noise spectrum in the frequency domain
  for channels attached to the longest wires from the three planes.
  The data is taken from before (dashed) and after (solid) the noise
  mitigating hardware upgrades were performed.  The plot zooms in on
  the frequency range of greatest interest.  Note that the HV drift
  system was turned on (at \SI{-70}{\kV}) at the time this data was
  taken, so minimum ionizing particle signals are included.  The
  low-frequency (\SIrange{10}{30}{\kHz}) noise and narrow-band
  (\SI{36}{\kHz}, \SI{108}{\kHz}, etc.) noise types are significantly 
  reduced by the hardware upgrades.}
\label{FFTcomp}
\end{figure}

To reduce the the low frequency coherent noise coming from the voltage
regulators, new service boards were installed that use
a different voltage regulator model (ADP7159). In addition, to reduce the
harmonic noise from the cathode HV power supply, a second ``filter
pot'' (low-pass filter) was installed in the drift HV
system. A comparison of the spectral noise density of the two types of
voltage regulators, as well as the predicted frequency response of the
one-pot and two-pot drift HV configurations are illustrated in
figure~\ref{HardwareFreqResp-1} and~\ref{HardwareFreqResp-2}.  The results of the hardware upgrades are
shown in figure~\ref{FFTcomp}, comparing the frequency content of the noise
spectrum of the longest wires on each wire plane before and after
the hardware noise mitigation. Both types of
excess noise are dramatically reduced after the upgardes in hardware were
implemented.

Table~\ref{tab:ENCcomp} summarizes the ENC associated with the
pedestal regions of waveforms from the longest wires from each of the
three planes, both before and after the hardware noise mitigation and
also after subsequent offline noise filtering.  The PSNR metrics
evaluated on raw data after the hardware upgrades alone and as
described above, are 18.1, 13.1 and 33.4 for the U, V, and Y planes,
respectively.  The hardware upgrades provided a noise suppression of
about a factor of 2 for the collection plane channels, and about a
factor of 3 for the induction planes.

It is important to note that there remains excess noise even after the
hardware upgrades, that is largely due to the incomplete mitigation of
the low frequency coherent noise from the voltage regulators. The
subsequent application of the offline noise filter provides
the needed additional noise suppression of 10-20\%.  In order to
achieve the lowest possible noise levels for TPC data, offline noise
filtering is still required.

\begin{table}[tb]
\center
\caption[ENC Comparison]{Summary of ENC for the longest wires on
    each plane.  Results for the raw data before and after the 
    hardware upgrades are raw data without any offline noise filtering.  Also
    shown are the results when offline noise filtering is applied to the
    data after hardware fixes.}
     \vspace{2mm}
  \label{tab:ENCcomp}
\begin{tabular}{|l|c|c|c|}
\hline
\bf{Waveform Type} & \bf{U Plane ENC} & \bf{V Plane ENC} & \bf{Y Plane ENC} \\ \hline
Before Hardware Fix & 1570 & 1340 & 640 \\
After Hardware Fix & 480 & 490 & 350 \\
Subsequent Offline Filter & 400 & 380 & 300 \\
\hline
\end{tabular}
\end{table}

\section{Conclusion and outlook}
\label{sec:summary}

In this paper, various sources of noise in the MicroBooNE TPC
are described. The noise is characterized in both the frequency and time
domains, as well as by wire length and channel number. The two largest sources
of excess noise have been identified as originating from the TPC drift
high-voltage power supply and the low-voltage regulators for the
front-end ASICs. A burst of significant noise at frequencies of around 
\SI{900}{\kHz} occurs occasionally in one section of the TPC, 
but its exact source 
has  yet to be identified. In addition, periodic saturation of some of the
ASICs has been observed and attributed to large current
induced by wire motion through the electric field caused by fluid flow.

Several offline noise filtering techniques have been developed to
remove most of the excess TPC noise. Further hardware solutions have
been implemented to address remaining sources of noise. The noise level
(ENC) after the offline noise filtering on the data before the hardware upgrades 
is in general below 400
$e^-$ for \SI{\approx90}{\percent} of MicroBooNE channels.  
There are 224 channels in the first induction U plane associated with 
an incorrect \SI{1}{\us}
peaking time setting that exhibit slightly higher noise. 
The residual noise is consistent 
with the cold electronics design expectations~\cite{asic} and test-stand
measurements and is found to be stable with time. The residual noise
levels are significantly lower than those in previous experiments utilizing 
warm front-end electronics and significantly improve the performance of 
the induction wire planes and the imaging capabilities of the TPC. 

Ten percent of the channels are unusable for the physics
analysis. The impact of these unusable channels depends on the
reconstruction strategy. 

The experience accumulated during the first year of data taking 
has proven to be critical in the optimization and operation 
of the MicroBooNE TPC and is useful in informing future LArTPC efforts.
In particular, two recent hardware upgrades have significantly 
reduced excess noise from the cathode HV power supply and the LV regulator. 
The resulting noise levels after the upgrades are 3.1, 2.5, and 1.8 times better for the 
U, V, and Y wire planes, respectively. With the observation of ASIC saturation, 
a new generation of ASICs for the future Short-Baseline Neutrino (SBN)~\cite{Antonello:2015lea} program and Deep Underground Neutrino Experiment (DUNE)~\cite{Acciarri:2016crz} has an added 
capability of setting \SI{1}{\nano\ampere} and \SI{5}{\nano\ampere} input bias currents. Spacers have also been 
added to support wires to reduce wire vibrations in the DUNE and SBND TPCs.
These spacers are also expected to reduce the impact of potential shorts that would be present in the case of a loose wire. 
With the observation of misconfigured channels, the ASIC design now 
implements more robust electrostatic discharge protection 
to configuration pins. The design margin of the bandgap reference circuit was 
also increased in the new ASIC design to remove the ``start-up'' problem.
As a result of this work, MicroBooNE has made
critical contributions in the understanding of noise levels in LArTPCs and in further mitigating such noise sources with offline noise filtering and future design improvements.


\begin{acknowledgments}
This material is based upon work supported by the following: the
U.S. Department of Energy, Office of Science, Offices of High Energy
Physics and Nuclear Physics; the U.S. National Science Foundation; the
Swiss National Science Foundation; the Science and Technology
Facilities Council of the United Kingdom; and The Royal Society
(United Kingdom). Additional support for the laser calibration system
and cosmic ray tagger was provided by the Albert Einstein Center for
Fundamental Physics. Fermilab is operated by Fermi Research Alliance,
LLC under Contract No. DE-AC02-07CH11359 with the United States
Department of Energy.
\end{acknowledgments}




\bibliographystyle{JHEP}

\bibliography{Noise}




\end{document}